\documentstyle[11pt,epsf]{article}

\newcommand{\newsection}{ \setcounter{equation}{0} \section}

\newcommand{\beq}{\begin{equation}} \newcommand{\eeq}{\end{equation}}
\newcommand{\bea}{\begin{eqnarray}} \newcommand{\eea}{\end{eqnarray}}

  \newcommand
{\Romannumeral}[1]{\uppercase\expandafter{\romannumeral#1}}

\newcommand{\be}{\begin{enumerate}} \newcommand{\ee}{\end{enumerate}}
\newcommand{\bi}{\begin{itemize}} \newcommand{\ei}{\end{itemize}}
\newcommand{\ba}{\begin{array}} \newcommand{\ea}{\end{array}}
\newcommand{\bc}{\begin{center}} \newcommand{\ec}{\end{center}}
\newcommand{\bt}{\begin{tabular}} \newcommand{\et}{\end{tabular}}

\def\lsim{\mathrel{\rlap{\lower4pt\hbox{\hskip1pt$\sim$}}
    \raise1pt\hbox{$<$}}}           
\def\gsim{\mathrel{\rlap{\lower4pt\hbox{\hskip1pt$\sim$}}
    \raise1pt\hbox{$>$}}}           

\newcommand{\tr}{\mathop{\rm tr}}           
\newcommand{\half}{\textstyle {1\over2} \displaystyle}    
\newcommand{\quarter}{\textstyle {1\over4} \displaystyle} 
\newcommand{\Dslash}{{\hbox{D}\kern-0.6em\raise0.15ex\hbox{/}}} 

\renewcommand{\a}{\alpha}
\renewcommand{\b}{\beta}

\renewcommand{\et}{\eta}

\newcommand{\e}{\epsilon}

\renewcommand{\l}{\lambda}
\newcommand{\m}{\mu}

\begin{document}

\setlength{\oddsidemargin}{0cm} \setlength{\baselineskip}{7mm}

\input epsf

\begin{normalsize}\begin{flushright}
     DAMTP-2005-109 \\
November 2005 \\
\end{flushright}\end{normalsize}

\begin{center}
  
\vspace{5pt}

{\Large \bf Quantum Gravity in Large Dimensions }

\vspace{10pt}

\vspace{10pt}

{\sl Herbert W. Hamber}
$^{}$\footnote{e-mail address : HHamber@uci.edu} \\
Department of Physics and Astronomy \\
University of California \\
Irvine, CA 92697-4575, USA \\

\vspace{5pt}

and

\vspace{5pt}

{\sl Ruth M. Williams}
$^{}$\footnote{e-mail address : R.M.Williams@damtp.cam.ac.uk} \\
Girton College, Cambridge CB3 0JG, and   \\
Department of Applied Mathematics and Theoretical Physics \\
Centre for Mathematical Sciences \\
Wilberforce Road, Cambridge CB3 0WA, United Kingdom.
\\

\end{center}

\begin{center} {\bf ABSTRACT } \end{center}

\noindent

Quantum gravity is investigated in the limit of a large number of
space-time dimensions,
using as an ultraviolet regularization the simplicial lattice path 
integral formulation.
In the weak field limit the appropriate expansion parameter is 
determined to be $1/d$.
For the case of a simplicial lattice dual to a hypercube, 
the critical point is found at $k_c/\lambda=1/d$ (with $k=1/8 \pi G$)
separating a weak coupling from a strong coupling phase, and
with $2 d^2$ degenerate zero modes at $k_c$.
The strong coupling, large $G$, phase is then investigated
by analyzing the general structure of the strong coupling expansion in the
large $d$ limit.
Dominant contributions to the curvature correlation functions
are described by large closed random polygonal surfaces,
for which excluded volume effects can be neglected at large $d$,
and whose geometry we argue can be approximated by unconstrained random
surfaces in this limit.
In large dimensions the gravitational correlation length is then found to
behave as $\vert \log ( k_c - k ) \vert^{1/2}$, 
implying for the universal gravitational critical exponent the value $\nu=0$
at $d=\infty$.

\vfill



\pagestyle{empty}

\newpage

\pagestyle{plain}

\vskip 10pt
\newsection{Introduction}
\hspace*{\parindent}

The lack of perturbative renormalizability for quantum gravitation
in physical dimensions \cite{feylec,dewittlec,thooft,deser,sagnotti,vandeven}
has brought to the forefront the need
to develop field theoretic approximation schemes 
that do not rely on the assumption of weak gravitational fields, and 
which are sophisticated enough to deal with the rich
physical structure of non-renormalizable theories
\cite{wilson,wilseps,parisi,sym,nonren-wein}.
The hope is that more powerful covariant methods, better suited to
the non-perturbative regime, will eventually shed some light on
the elusive long distance properties of quantum gravitation, which could 
ultimately have a bearing
on a number of long standing and fundamental issues,
such as the short distance nature of space-time, the emergence
of the semiclassical limit
and the problem of large-scale quantum cosmology \cite{hawking,
harhawk,harcosm,cosm}.

Approaches based on the simplicial lattice formulation for gravity
\cite{regge,wheeler,rowi,lee,hw84,hartle,lesh,monte,critical},
the $2+\epsilon$ 
expansion \cite{wilseps,epsilon,epsilon1,epsilon2,epsilon3}
and approximate renormalization group methods based on Wilson's
momentum slicing technique \cite{litim,reuter} have suggested the
existence of a nontrivial ultraviolet fixed point in and around
four dimensions, separating a weakly coupled
(but physically un-attractive) phase from a strongly coupled one,
the latter phase being characterized by a finite invariant correlation length,
and close to smooth geometries at large distances.
Substantial uncertainties remain in each of the three approximation
methods mentioned above, both about the results
themselves and their relationship to each other, but also
regarding their ultimate physical significance and how they might relate
to physical gravitational phenomena, and both early and late time cosmology.
It would be clearly desirable if one could find a limit in the
quantum gravity case where non-perturbative aspects of the theory 
could be fully explored by covariant analytical means.
In the significantly simpler Yang-Mills case the evidence so far
is that the lattice is the only reliable non-perturbative method,
capable of producing reasonably unambiguous quantitative results,
within a controlled approximation based on the zero lattice spacing limit.
It will be this method that will be therefore the focus of our work.

In this paper we study a set of approximation methods based
on an expansion in the inverse number of dimensions.
Increasing the number of space-time dimensions above four only worsens
the renormalizability problem, which implies that the need for
a non-perturbative approach, such as the lattice one, becomes even more acute.
The so-called $1/d$ expansion was originally developed for
statistical mechanics systems, and later extended to the study
of quantum field theory, where it has since met with a number
of considerable successes, including an understanding of triviality
for scalar field theories above four dimensions (which,
incidentally, are not perturbatively renormalizable for any $d>4$).
The above expansion is known to be intimately tied up with the mean
field theory treatment of quantum mechanical systems, but not necessarily
equivalent to it (as was already noted in the gauge theory case),
and exploits the fact that in large dimensions each point is
typically surrounded by many neighbors, whose action
can then be either treated exactly, or included as some sort of local average.
For classical spin systems at finite temperature, the $1/d$ expansion
was originally developed in \cite{englert,fisher,abe} by examining
the structure of the high temperature expansion.

In many ways the $1/d$ expansion is similar to the very successful
$1/N$ expansion for statistical mechanics systems (the $O(N)$ vector
model being one thoroughly explored and well-understood 
example \cite{stanley,zinn}) and $SU(N)$ gauge theories,
where it leads to the planar diagram approximation \cite{thooft1,thooft2}
and the many phenomenological successes that follow from it.
In the gravitational case it is less obvious how to attach color
degrees (or any other internal degree) of freedom to the graviton,
so this particular avenue seems unfruitful at the moment.

In this paper we will study large-dimensional pure gravitation, 
without any matter fields, which could then be added at a later stage.
We recall here that for pure gravity in $d$ dimensions there are $d(d+1)/2$
independent components of the metric, and the same number of
algebraically independent components of the Ricci tensor. 
The contracted Bianchi identities then reduce the
count by $d$, and so does general coordinate invariance,
leaving $d(d+1)/2 - d - d = d(d-3)/2$ physical gravitational
degrees of freedom in $d$ dimensions.
As a result, the number of physical degrees of freedom of the gravitational
field grows rather rapidly (quadratically) with the number of dimensions.

The paper is organized as follows.
In Section 2 we discuss the machinery of the $1/d$ expansion for
the lattice theory of gravity based on Regge's simplicial construction.
The action simplifies considerably in the large $d$ limit, and
we are able to exhibit the location of the critical point
in the variable $k=1/8 \pi G$, at least in the weak field limit,
as well as the nature of the excitation spectrum around it. 
In Section 3 we follow a complementary route to the large $d$ limit,
where we perform a simultaneous $1/d$ and strong coupling (small $k$)
expansion. 
Since the strong coupling expansion for simplicial lattice gravity
has not been discussed before in the literature, we will present here
some general aspects of it.
We then show how the relevant (in the long distance limit)
critical behavior can be extracted from the strong coupling expansion
by analyzing the geometric structure of its dominant terms.
In Section 4 we provide some contact with results obtained in
the continuum in and above $d=4$, and compare and contrast with
what has been found in the previous two sections from the simplicial
lattice theory.
Appendix A contains a brief summary of the large $d$ limit 
for scalar lattice field theories, while
Appendix B discusses some results relevant to non-Abelian gauge fields
on the lattice.


\vskip 30pt
\newsection{Expansion in Inverse Powers of the Dimension}
\hspace*{\parindent}

Our first concern will be an approximate evaluation, in the
large $d$ limit, of an appropriately 
discretized form of the continuum Euclidean functional integral
for pure gravity without matter, which we write here as
\beq
Z_{cont} \; = \; \int \left [ d \, g_{\mu\nu} \right ] \; \exp \,
\left (  - \lambda \, \int d^d x \, \sqrt g \, + \, 
{ 1 \over 16 \pi G } \int d^d x \sqrt g \, R  \, \right ) \;\; .
\label{eq:zc}
\eeq
In the following we will therefore first address the key issue of 
precisely what type of terms in the discrete action, based on the simplicial
lattice formulation \cite{regge}, become dominant in this limit.

\subsection{General formulae in $d$ dimensions}

We will consider here a general simplicial lattice in $d$ dimensions, made out
of a collection of flat $d$-simplices glued together at their common faces 
so as to constitute a triangulation of a smooth continuum manifold,
such as the $d$-torus or the surface of a sphere.
Each simplex is endowed with $d+1$ vertices, and its geometry
is completely specified by assigning the lengths of its $d(d+1)/2$ edges.
We will label the vertices by $1, 2, 3, \dots d+1$ and denote the 
square edge lengths by $l _ {12}^2 = l _ {21}^2$, ... $l _ {1,d+1}^2 $.
The vertices of the simplex can be specified by a set of vectors 
${\bf v}_1 =0$, ${\bf v}_2$, ... ${\bf v}_{d+1}$.
The matrix
\beq
g_{ij} \; = \; < {\bf v}_{i+1} \vert {\bf v}_{j+1} > \;\; , 
\eeq
with $1 \leq i,j \leq d $, is positive definite,
and, in terms of the edge lengths
$l_{ij} \, = \, | {\bf v}_i - {\bf v}_ j | $, it is given by
\beq
g_{ij} \; = \; {1 \over 2} 
\left ( l_{1,i+1}^2 + l_{1,j+1}^2 - l_{i+1,j+1}^2 \right ) \;\; .
\label{eq:latmet}
\eeq
The volume of a $d$-simplex is then given by the $d$-dimensional
generalization of the well-known formula for a tetrahedron
\beq
V_d \; = \; {1 \over d ! }  \sqrt { \det  g_{ij} } \;\; .
\eeq
An equivalent form can be given in terms of a determinant
of a $(d+2) \times (d+2)$ matrix,
\beq
V_d \; = \; {(-1)^{d+1 \over 2 } \over d! \, 2^{d/2} } \,
\left|\matrix{
0      &    1     &    1     & \ldots \cr
1      &    0     & l_{12}^2 & \ldots \cr 
1      & l_{21}^2 &    0     & \ldots \cr
1      & l_{31}^2 & l_{32}^2 & \ldots \cr
\ldots &  \ldots  &  \ldots  & \ldots \cr
1      & l_{d+1,1}^2 & l_{d+1,2}^2 & \ldots \cr
}\right| ^{1/2}  .
\label{eq:vol}
\eeq
Then the dihedral angle in a $d$-dimensional simplex of volume
$V_d$, between faces of volume $V_{d-1}$ and $V_{d-1}^{'}$, is obtained from
\beq
\sin \theta_d \; = \; { d \over d-1 } \, 
{ V_d \, V_{d-2} \over V_{d-1} \, V_{d-1}^{'} } \;\; .
\eeq
In the equilateral case we record here the particularly simple result
for the volume of a simplex
\beq
V_d \; = \; { 1 \over d! } \sqrt{ d+1 \over 2^d } \;\; ,
\label{eq:vequilat}
\eeq
and for the dihedral angle
\beq
\cos \theta_d \; = \; { 1 \over d } \;\; .
\eeq
The $d$-dimensional Euclidean lattice action, involving cosmological
constant and scalar curvature terms, is then given by
\beq
I (l^2) \; = \; 
\lambda \, \sum \, V_d \; - \; k \, \sum \,  \delta_d \, V_{d-2} \;\; ,
\label{eq:latac}
\eeq
and appears in the partition function as
\beq
Z ( \lambda, \, k ) \; = \; \int [ d \, l^2 ] \, \exp \left ( - I(l^2)
\right ) \;\; .
\label{eq:zdef}  
\eeq

\subsection{Weak field expansion}

The above formulae for volumes and angles are quite complicated
in the general case, and therefore of limited use in large dimensions.
The next step consists in expanding them out in 
terms of small edge length variations,
\beq
l_{ij}^2 \; = \; l_{ij}^{(0)\, 2} \; + \; \delta \, l_{ij}^2 \;\; .
\eeq
We will set for convenience from now on 
$ \delta \, l_{ij}^2 \, = \, \e_{ij} $.
Unless stated otherwise, we will be considering the expansion about
the equilateral case, and set $l_{ij}^{(0)} = 1$
(we will later relax this last restriction).
Furthermore one has the well known expansion for determinants
\bea
\det ( 1 + M ) \, & = & \, e^{ \tr \ln (1+M) }  
\nonumber \\
& = & \, 1 \, + \, \tr M \, + \, 
{1 \over 2!} \left [ ( \tr M )^2 \, - \, \tr M^2 \right ]
\, + \, 
{1 \over 3!} \left [ ( \tr M )^3 \, - \, 3 \, \tr M \, \tr M^2 
\, + \, 2 \, \tr M^3 \right ] \, + \, \dots
\nonumber \\
\eea
One can then re-write the expression in Eq.~(\ref{eq:vol}) for the volume of
a $d$-simplex as 
\beq
V_d \; = \; {(-1)^{d+1 \over 2 } \over d! \, 2^{d/2} } \, \sqrt{ \det
  M_d } \;\; ,
\eeq
and expanding out to quadratic order one finds
\beq
\sqrt{ - \det M_2 } \; = \; \sqrt{3} \, + \, 
{ 1 \over \sqrt{3} } \, \e_{12} \, + \, \dots 
\, + \, { 2 \over 3 \sqrt{3} } \, \e_{12} \, \e_{13}
\, + \, \dots 
\, - \, { 2 \over 3 \sqrt{3} } \, \e_{12}^2 \, + \, \dots \;\; , 
\eeq
\beq
\sqrt{ \det M_3 } \; = \; \sqrt{4} \, + \, 
{ 1 \over \sqrt{4} } \, \e_{12} \, + \, \dots 
\, + \, { 3 \over 4 \, \sqrt{4} } \, \e_{12} \, \e_{13}
\, + \, \dots 
\, - \, { 1 \over 4 \, \sqrt{4} } \, \e_{12} \, \e_{34}
\, + \, \dots 
\, - \, { 9 \over 2 \cdot 4 \sqrt{4} } \, \e_{12}^2 \, + \, \dots 
\eeq
and for general $d$
\beq
{ 1 \over \sqrt{d+1} } \, \sqrt{ \pm \, \det M_d } \; = \; 
1 \, + \, 
{ 1 \over d + 1 } \, \e_{12} \, + \, 
{ d \over (d+1)^2 } \, \e_{12} \, \e_{13}
\, - \, { 1 \over (d+1)^2 } \, \e_{12} \, \e_{34}
\, - \, { d^2 \over 2 \, (d+1)^2 } \, \e_{12}^2 \, + \, \dots 
\, + \, O( \e^2 ) .
\eeq
For large $d$ the last expression simplifies to
\beq
{ 1 \over \sqrt{d+1} } \, \sqrt{ \pm \, \det M_d } \; = \; 
1 \, + \, 
{ 1 \over d } \, ( \e_{12} \, + \, \dots ) 
\, + \, 
{ 1 \over d } \, ( \e_{12} \, \e_{13} \, + \, \dots ) 
\, - \, 
{ 1 \over d^2 } \, ( \e_{12} \, \e_{34} \, + \, \dots ) 
\, - \, 
{ 1 \over 2 } \, ( \e_{12}^2 \, + \, \dots ) \, + \, O( \e^3 ) . 
\label{eq:vold}
\eeq
Here the terms $\e_{12} \, \e_{13} $ refer to two edges
sharing a common vertex, whereas the terms $\e_{12} \, \e_{34}$
denote terms with opposite edges, not sharing a common vertex.

As a result, the volume term appearing in the $d$-dimensional
Euclidean lattice action of Eq.~(\ref{eq:latac}), becomes
\beq
V_d \; \mathrel{\mathop\sim_{ d \rightarrow \infty }} \;
{ \sqrt{d} \over d! \, 2^{d/2} } \, \left \{ 
1 \, + \, {1 \over d } ( \e_{12} \, + \, \dots ) \, + \, 
{1 \over d }   ( \e_{12} \, \e_{13} \, + \, \dots ) \, - \,
{1 \over d^2 } ( \e_{12} \, \e_{34} \, + \, \dots ) \, - \,
{1 \over 2 } ( \e_{12}^2 \, + \, \dots ) \, + \, \dots \right \} ,
\eeq
or, equivalently, ordering the terms in powers of $1/d$,
\beq
V_d \; \mathrel{\mathop\sim_{ d \rightarrow \infty }} \;
{ \sqrt{d} \over d! \, 2^{d/2} } \, \left \{ 
1 \, - \, 
{1 \over 2 } \, \e_{12}^2 \, + \, \dots \, + \,
{1 \over d } \, 
( \e_{12} \, + \, \dots \, + \, \e_{12} \, \e_{13} 
\, + \, \dots ) \, + \, O ( {1 \over d^2 } )  \right \} \;\; .
\label{eq:volex}
\eeq
To leading order, it involves a lattice sum over all
squared edge length deviations.
Note that the terms linear in $\e$ (the so called tadpole terms
in the continuum), which would have required a shift in the ground state
value of $\e$ for a non-vanishing cosmological constant $\lambda$,
vanish to leading order in $1/d$.
The full volume term $\lambda \sum V_d $ appearing in the action can then
be easily written down using the above expressions.

Next one needs to expand the dihedral angle. 
In the equilateral case one has for the dihedral angle
\beq
\theta_d \; = \; \arcsin { \sqrt{d^2 -1} \over d } \; 
\mathrel{\mathop\sim_{ d \rightarrow \infty } } \;
{ \pi \over 2 } \, - \, { 1 \over d } \, - \, { 1 \over 6 \, d^3 } 
\, + \, \dots \;\; ,
\label{eq:arcsin}
\eeq
which will require {\it four} simplices to meet on a hinge, to give
a deficit angle of $ 2 \pi - 4 \times {\pi \over 2 } \approx 0 $ in large
dimensions. 
One notes that in large dimensions the simplices look locally
(i.e. at a vertex) more like hypercubes.
Several $d$-dimensional simplices will meet on a $(d-2)$-dimensional
hinge, sharing a common face of dimension $d-1$ between adjacent simplices.
Each simplex has $(d-2)(d-1)/2$ edges ``on'' the hinge, 
some more edges are then situated on the two ``interfaces'' between
neighboring simplices meeting at the hinge, and finally one edge 
lies ``opposite'' to the hinge in question.
In two dimensions one finds for the dihedral angle at vertex $1$,
to quadratic order,
\bea
\theta_2 \; & = & \; { \pi \over 3 } 
\, - \, { 1 \over 2 \, \sqrt{3} } \, ( \e_{12} \, + \, \e_{13} \, ) 
\, + \, { 1 \over \sqrt{3} } \, \e_{23}
\nonumber \\
\, & + & \, { 1 \over 12 \, \sqrt{3} } \, ( \e_{12}^2 \, + \, \e_{13}^2 \, ) 
\, + \, { 2 \over 3 \, \sqrt{3} } \, \e_{12} \, \e_{13}
\, - \, { 1 \over 3 \, \sqrt{3} } \, ( \e_{12} \, + \, \e_{13} \, ) \, \e_{23}
\, - \, { 1 \over 6 \, \sqrt{3} } \, \e_{23}^2 \;\; ,
\eea
whereas in three dimensions one has for the dihedral
angle at edge $12$, to the same order,
\bea
\theta_3 \; & = & \; \arcsin { 2 \sqrt{2} \over 3 } 
\, + \, { 1 \over 3 \, \sqrt{2} } \, \e_{12} 
\, - \, { 1 \over 3 \, \sqrt{2} } \, ( \e_{13} \, + \, \e_{14} \, + \, \e_{23} \, + \, \e_{24} \, ) 
\, + \, { 1 \over \sqrt{2} } \, \e_{34} 
\nonumber \\
\, && + \, { 7 \over 72 \, \sqrt{2} } \, \e_{12}^2
\, - \, { 1 \over 72 \, \sqrt{2} } \, ( \e_{13}^2 \, + \, \e_{14}^2 \, + \, \e_{23}^2 \, + \, \e_{24}^2 \, ) 
\, - \, { 7 \over 36 \, \sqrt{2} } \,  \e_{12} \, ( \e_{13} \, + \, \e_{14} \, + \, \e_{23} \, + \, \e_{24} )
\nonumber \\
\, && - \, { 1 \over 4 \, \sqrt{2} } \, ( \e_{13} \, \e_{24} \, + \, \e_{14} \, \e_{23} )
\, + \, { 3 \over 4 \, \sqrt{2} } \, ( \e_{13} \, \e_{14} \, + \, \e_{23} \, \e_{24} )
\, + \, { 11 \over 36 \, \sqrt{2} } \, ( \e_{13} \, \e_{23} \, + \, \e_{14} \, \e_{24} )
\nonumber \\
\, && + \, { 1 \over 4 \, \sqrt{2} } \, \e_{12}  \, \e_{34}
\, - \, { 1 \over 4 \, \sqrt{2} } \, 
( \e_{13} \, + \, \e_{14} \, + \, \e_{23} \, + \, \e_{24} ) \, \e_{34}
\, - \, { 1 \over 8 \, \sqrt{2} } \, \e_{34}^2 \;\; .
\eea
In the general $d$-dimensional case the expansion coefficients for the
dihedral angle at the hinge labeled by $1, \, 2, \dots d-1$
are given by the following expressions (as well as their
large d limit)
\bea
{ 2 \over d \sqrt{d^2 -1} } \, \e_{12} \; & \rightarrow & \; 
{ 2 \over d^2 } \, \e_{12} 
\nonumber \\
- { d-1 \over d \sqrt{d^2 -1} } \, \e_{1,d} \; & \rightarrow & \; 
- { 1 \over d } \, \e_{1,d} 
\nonumber \\
{ d-1 \over \sqrt{d^2 -1} } \, \e_{d,d+1} \; & \rightarrow & \; 
\, \e_{d,d+1} 
\nonumber \\
{ 2 ( d^3 -2 d^2 -d+1 ) \over d^2 (d^2 -1)^{3/2} } \, \e_{12}^2 \; & \rightarrow & \; 
{ 2 \over d^2 } \, \e_{12}^2 
\nonumber \\
- { ( d^2 - 2 d - 2 ) (d-1)^2 \over 2 d^2 (d^2 -1)^{3/2} } \, \e_{1,d}^2 \; & \rightarrow  & \; 
- { 1 \over 2 d } \, \e_{1,d}^2 
\nonumber \\
- { (d-1)^2 \over 2 (d^2 -1)^{3/2} } \, \e_{d,d+1}^2 \; & \rightarrow & \; 
- { 1 \over 2 d } \, \e_{d,d+1}^2 
\nonumber \\
{ 2 ( d^3 -4 d^2 -d+2 ) \over d^2 (d^2 -1)^{3/2} } \, 
\e_{12} \, \e_{13}  \; & \rightarrow & \; 
{ 2 \over d^2 } \, \e_{12} \, \e_{13}
\nonumber \\
- { 4 ( 2 d^2 - 1 ) \over d^2 (d^2 -1)^{3/2} } \, 
\e_{12} \, \e_{34}  \; & \rightarrow & \; 
- { 8 \over d^3 } \, \e_{12} \, \e_{34}
\nonumber \\
- { (d-1) ( d^3 -2 d^2 +d+2 ) \over d^2 (d^2 -1)^{3/2} } \, 
\e_{12} \, \e_{1,d+1}  \; & \rightarrow & \; 
- { 1 \over d } \, \e_{12} \, \e_{1,d+1}
\nonumber \\
{ 2 ( d^2  -d -1 ) \over d^2 (d+1) \sqrt{d^2 -1} } \, 
\e_{34} \, \e_{1,d+1}  \; & \rightarrow & \; 
{ 2 \over d^2 } \, \e_{34} \, \e_{1,d+1}
\nonumber \\
{ 2 \over (d+1) \sqrt{d^2 -1} } \, 
\e_{12} \, \e_{d,d+1}  \; & \rightarrow & \; 
\, {2 \over d^2 } \, \e_{12} \, \e_{d,d+1}
\nonumber \\
{ d (d-1) \over (d+1) \sqrt{d^2 -1} } \, 
\e_{1,d} \, \e_{1,d+1}  \; & \rightarrow & \; 
\e_{1,d} \, \e_{1,d+1}
\nonumber \\
{ (d-1) (3d+2 ) \over d^2 (d+1) \sqrt{d^2 -1} } \, 
\e_{1,d} \, \e_{3,d}  \; & \rightarrow & \; 
{ 3 \over d^2 } \, \e_{1,d} \, \e_{3,d}
\nonumber \\
- { (d-1) \over (d+1) \sqrt{d^2 -1} } \, 
\e_{1,d} \, \e_{3,d+1}  \; & \rightarrow & \; 
- { 1 \over d } \, \e_{1,d} \, \e_{3,d+1}
\nonumber \\
- { (d-1) \over (d+1) \sqrt{d^2 -1} } \, 
\e_{1,d} \, \e_{d,d+1}  \; & \rightarrow & \; 
- { 1 \over d } \, \e_{1,d} \, \e_{d,d+1} \;\; .
\nonumber \\
\eea
In the large $d$ limit one then obtains, to leading order
\bea
\theta_d & \; \mathrel{\mathop\sim_{ d \rightarrow \infty }} \; &
\arcsin { \sqrt{d^2 -1} \over d } \, + \,
\e_{d,d+1} \, + \, \e_{1,d} \, \e_{1,d+1} \, + \, \dots 
\nonumber \\
\, && + \, {1 \over d } \, \left ( 
- \, \e_{1,d} \, + \, \dots \, - \, 
{ 1 \over 2 } \, \e_{1,d}^2 \, + \, \dots \, - \, 
{ 1 \over 2 } \, \e_{d,d+1}^2 
\, - \, \e_{12} \, \e_{1,d+1} 
\, - \, \e_{1,d} \, \e_{3,d+1} 
\, - \, \e_{1,d} \, \e_{d,d+1}  \, + \, \dots \right )
\nonumber \\
&& \, + \, O ( {1 \over d^2 } ) \;\; .
\label{eq:dihedr}
\eea
To evaluate the curvature term $- k \sum \delta_d V_{d-2} $
appearing in the gravitational lattice action one needs the
hinge volume $V_{d-2}$, which is easily obtained from Eq.~(\ref{eq:volex}), 
by reducing $d \rightarrow d-2$,
\beq
V_{d-2} \; \mathrel{\mathop\sim_{ d \rightarrow \infty }} \;
{ 2 d^{3/2} (d-1) \over d! \, 2^{d/2} } \, \left \{ 
1 \, - \, 
{1 \over 2 } \e_{12}^2 \, + \, \dots \, + \,
{1 \over d } ( \e_{12} \, + \, \dots \, + \, \e_{12} \, \e_{13} \, +
\, \dots ) \, + \, O ( {1 \over d^2 } )  \right \} \;\; ,
\label{eq:hingevol}
\eeq
whereas the deficit angle $\delta$ is given by
\beq
\delta_d \; = \; 2 \, \pi - \sum_{\rm simplices} \theta_d 
\; = \; 2 \, \pi - \sum_{\rm simplices} \left \{ \, \arcsin {
  \sqrt{d^2-1} \over d } \, + \, \dots \right \} \;\; ,
\eeq
with the expansion of the $arcsin$ function given in Eq.~(\ref{eq:arcsin}).

\subsection{Evaluation of the lattice action}

We now specialize to the case where four simplices meet at a hinge.
When expanded out in terms of the $\e$'s one obtains for the
deficit angle
\bea
\delta_d \; & = & \; 2 \, \pi - 4 \cdot {\pi \over 2} \, + \,
\sum_{\rm simplices} { 1 \over d } \, - \, \e_{d,d+1} \, + \, \dots 
\, - \, \e_{1,d} \, \e_{1,d+1} \, + \, \dots 
\nonumber \\
\, && - \, {1 \over d } \, \left ( 
- \, \e_{1,d} \, - \, { 1 \over 2 } \, \e_{1,d}^2
\, - \, { 1 \over 2} \, \e_{d,d+1}^2 \, - \, 
\e_{12} \, \e_{1,d+1} \, - \, 
\e_{1,d} \, \e_{3,d+1} \, - \, 
\e_{1,d} \, \e_{d,d+1} \, + \, \dots \right ) 
\, + \, O( {1 \over d^2} ) .
\nonumber \\
\label{eq:deficit}
\eea
The action contribution involving the deficit angle is then, for a single hinge,
\bea
- \, k \, \delta_d \, V_{d-2} \; & = & \;
( - \, k ) \, 
{ 2 \, d^{3/2} \, (d-1) \over d! \, 2^{d/2} } \, \left \{ 
1 \, - \, 
{1 \over 2 } \e_{12}^2 \, + \, \dots \,  \right \}
\, \left \{ {4 \over d} \, + \, \dots \, - \, 
\e_{d,d+1} \, + \, \dots \, - \,
\e_{1,d} \, \e_{1,d+1} \, + \, \dots \, \right \}
\nonumber \\
\, & = & \; ( - \, k ) \, 
{ 2 \, d^{3/2} \, (d-1) \over d! \, 2^{d/2} } \, \left (
\, - \, \e_{d,d+1} \, + \, \dots \, - \, 
\e_{1,d} \, \e_{1,d+1} \, + \, \dots \, \right ) \;\; .
\label{eq:curvterm}
\eea
It involves two types of terms: one linear in the (single) edge
opposite to the hinge, as well as a term involving a product
of two distinct edges, connecting any hinge vertex to the two vertices
opposite to the given hinge.
Since there are four simplices meeting on one hinge, one will
have 4 terms of the first type, and $4(d-1)$ terms of the second type.
Combining the cosmological constant and the curvature contributions
one then obtains
\beq
{ \sqrt{d} \over d! \, 2^{d/2} } \, \left [
\, \lambda \, \left ( 1 \, - \, {1 \over 2 } \, \e_{12}^2 \, + \, 
{\sigma \over 4} \, \e_{12}^4 \, + \, \dots \,  \right )
\, - \, k \cdot 2 \, d \, (d-1) \, \left ( 
\, - \, \e_{d,d+1} \, + \, \dots \, - \,
\e_{1,d} \, \e_{1,d+1} \, + \, \dots \, \right ) \, \right ] .
\eeq
The first term in the above expression refers to a single simplex,
the second one to a single hinge. 
To obtain the total action, a sum over all simplices, resp. hinges, has
still to be performed.

We have also added a term $\sigma \e^4 $ in order to impose
a cutoff at large edge lengths $ | \e | $.
The justification for this choice comes from the fact that numerical
simulations show convincingly that very large, as well as very small,
edge lengths are exponentially suppressed by the lattice gravitational measure,
and in particular by a non-trivial interplay between the $\lambda$ term
and the generalized triangle inequalities \cite{hw84,hartle,critical,cms1,det}
(as such, $\sigma$ is not really a parameter that one is allowed to vary,
and should rather be fixed to some suitable numeric value).
Dropping the irrelevant constant term and summing over edges
one obtains for the total action 
$\lambda \sum V_d - k \sum \delta_d \, V_{d-2}$
in the large $d$ limit
\beq
\lambda \, \left ( \, - \, {1 \over 2 } \, \sum \e_{ij}^2 \, + \,
{\sigma \over 4} \, \sum \e_{ij}^4 \, \right )
\, - \, 2 \, k \, d^2 \, \left ( 
\, - \, \sum \e_{jk} \, - \, \sum \e_{ij} \, \e_{ik} \, 
\right ) \;\; ,
\label{eq:totact}
\eeq
up to an overall multiplicative factor $ \sqrt{d} / d! \, 2^{d/2} $,
which will play no essential role in the following.
The $ \e_{ij} \, \e_{ik} $ coupling terms in the expression above
can always of course be re-written in terms of finite differences,
\beq
\e_{ij} \, \e_{ik} \; = \; - {1 \over 2} \,
\left ( \e_{ij} \, - \, \e_{ik} \right )^2 
\, + \, {1 \over 2} \, \e_{ij}^2 \, + \, {1 \over 2} \, \e_{ik}^2 ,
\eeq
and for smooth enough fields the first term on the r.h.s can be
regarded as a discrete approximation to a derivative.

From the action in of Eq.~(\ref{eq:totact}), one notices
that its form leads naturally to a first rough estimate for the critical point,
defined as the point where the competing $\lambda$ and curvature terms
achieve comparable magnitudes, namely $k_c \sim \lambda / d^2 $.
This results will be further improved below when we perform an explicit
calculation, which takes into account the actual number of
neighbors for each point, given a specific choice of lattice and its
associated coordination number (see Eq.~(\ref{eq:kcd1})).

\subsection{Action for the surface of the cross polytope}

The next step involves the choice of a specific lattice on which the
action is then evaluated. 
One possibility would be the hypercubic lattice, divided into simplices
as originally discussed in \cite{rowi}.
This type of lattice has $ 2^d -1$ edges emanating from each site
in $d$ dimensions
\footnote{
Which should be compared to the $\sim d^2 / 2 $ transverse-traceless
degrees of freedom of the continuum gravitational field in $d$ dimensions.
The exponential growth for this particular lattice implies the existence
of many redundant degrees of freedom in the large $d$ limit.
Amusingly, it is reminiscent of the Dirac spinor case, for which the
number of degrees of freedom is also exponential, $\sim 2^{d/2}$ for large $d$.
}.
Here we will evaluate the above action for the cross polytope $\beta_{d+1}$.
The cross polytope $\beta_n$
is the regular polytope in $n$ dimensions corresponding
to the convex hull of the points formed by permuting the coordinates 
$(\pm 1, 0, 0, ..., 0)$, and has therefore $2n$ vertices.
It is named so because its  vertices are located equidistant
from the origin, along the Cartesian axes in $n$-space.
The cross polytope in $n$ dimensions
is bounded by $2^n$ $(n-1)$-simplices, has $2n$ vertices
and $2n(n-1)$ edges.
In three dimensions, it represents the convex hull of the
octahedron, while in four dimensions the cross polytope  is the 16-cell
\cite{coxeter}.
In the general case it is dual to a hypercube in $n$ dimensions, with
the `dual' of a regular polytope being another regular
polytope having one vertex in the center of each cell of the polytope 
one started with.

\begin{center}
\epsfysize=8.0cm
\epsfbox{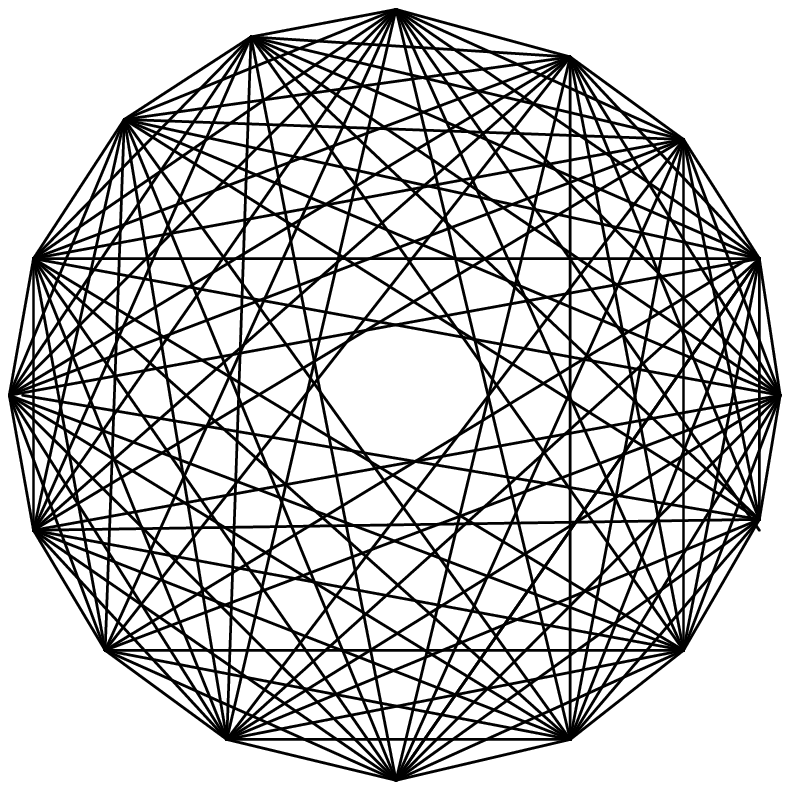}
\end{center}

\noindent{\small Fig 1. Cross polytope $\beta_n$ with $n=8$ and $2n=16$
vertices, whose surface can be used to define a simplicial manifold of
dimension $d=n-1=7$. For general $d$, the cross polytope $\beta_{d+1}$
will have $2(d+1)$ vertices, connected to each other by $2d(d+1)$ edges. 
\medskip}

When we consider the surface of the cross polytope in $d+1$ dimensions,
we have an object of dimension $n-1=d$, which 
corresponds to a triangulated manifold with no boundary, homeomorphic
to the sphere (as an example, see Fig. 1).
The deficit angle is given to leading order by
\beq
\delta_d \; = \; 0 \, + \,  { 4 \over d } \, - \, \left (
\e_{d,d+1} \, + \, {\rm 3 \, terms} \, + \, 
\e_{1,d} \, \e_{1,d+1} \, + \, \dots \right ) \, + \, 
O( 1/d^2, \e / d , \e^2 / d )
\eeq
and therefore close to flat in the large $d$ limit.
Indeed if the choice of triangulation is such that the deficit angle is not
close to zero, then the discrete model leads to an average curvature
whose magnitude is comparable to the lattice spacing or ultraviolet cutoff,
which from a physical point of view does not seem very attractive:
one obtains a spacetime with curvature radius comparable to the
Planck length.
In addition, the small fluctuation excitation spectrum for such
strongly curved lattices looks disturbingly different from what one
would expect in the continuum for transverse-traceless modes \cite{hartle}.

When evaluated on such a manifold the lattice action becomes
\beq
{ \sqrt{d} \, 2^{d/2} \over d! } \, 2 \,
\left ( \lambda \, - \, k \, d^3 \right ) \, \left [ 
1 \, - \, {1 \over 8} \, \sum \e_{ij}^2 \, + \, 
{ 1 \over d } \left ( { 1 \over 4 } \sum \e_{ij} \, + \, 
{ 1 \over 8 } \sum \e_{ij} \, \e_{ik} \right ) \, + \, O(1/d^2) 
\right ] \;\; .
\label{eq:beta}
\eeq
Dropping the $1/d$ correction one obtains to leading order
\beq
{ \sqrt{d} \, 2^{d/2} \over d! } \, 2 \,
\left ( \lambda \, - \, k \, d^3 \right ) \,  
\left ( 1 \, - \, {1 \over 8} \, \sum \e_{ij}^2 \, + \, \dots \right )
\label{eq:beta1}
\eeq
and, up to the irrelevant constant term and an overall multiplicative
factor, which can be absorbed into a re-scaling of the $\e$'s,
the action reduces to the simple form
\beq
- \, { 1 \over 2 } \, \left ( \lambda \, - \, k \, d^3 \right ) \, 
\sum \e_{ij}^2 \;\; .
\label{eq:beta2}
\eeq
Since there are $2 d (d+1)$ edges in the cross polytope, one finds
therefore that, at the critical point $k d^3 = \lambda$, the
quadratic form in $\e$, defined by the above action,
develops $2 d (d+1) \sim 2 d^2 $ zero eigenvalues
\footnote{
This result is quite close to the $d^2/2$ zero eigenvalues expected
in the continuum for large $d$, with the factor of four discrepancy
presumably attributed to an underlying intrinsic ambiguity that
arises when trying to identify lattice points with continuum points.
}.

It is worth noting here 
that the competing curvature ($k$) and cosmological constant 
($\lambda$) terms will have comparable magnitude when
\beq
k_c \; = \; { \lambda \, l_0^4 \over \, d^3 \, l_0^2 } \;\; .
\label{eq:kc4}
\eeq
Here we have further allowed for the possibility that the
average lattice spacing $l_0 = \langle l^2 \rangle^{1/2}$ is not equal to one
(in other words, we have restored the appropriate overall scale
for the average edge length, which is in fact largely determined by the
value of $\lambda$).
This then gives for $\lambda=1$ (using the large-$d$ expression for
the average lattice spacing $l_0$, obtained later in this section in
Eq.~(\ref{eq:l2d})), the estimate 
$k_c = \sqrt{3} / (16 \cdot 5^{1/4} ) = 0.0724$ in $d=4$,
to be compared with $k_c = 0.0636(11)$ obtained in \cite{critical}
by direct numerical simulation in four dimensions.
Even in $d=3$ one finds for $\lambda=1$, from Eqs.~(\ref{eq:kc4}) 
and (\ref{eq:l2d}), $k_c= 2^{5/3} / 27 = 0.118$, to be compared with 
$k_c=0.112(5)$ obtained in \cite{hw3d} by direct numerical simulation.
Again, the dependence of $k_c$ on inverse powers of $d$ is not surprising, 
as fluctuations, which are
stronger in smaller dimensions, will require an increasingly
larger value of the coupling $k$ to make the transition happen
in small dimensions.

The average lattice spacing $l_0$ is easily estimated from the
following argument. 
The volume of a general equilateral simplex is given by
Eq.~(\ref{eq:vequilat}), multiplied by an additional factor of $l_0^d$.
In the limit of small $k$ the average volume of a simplex
is largely determined by the cosmological term, 
and can therefore be computed from
\beq
< V > \; = \; - \, {\partial \over \partial \lambda } \,
\log \, \int [dl^2] \, e^{- \lambda V (l^2) } \;\; ,
\eeq
with $V(l^2) = ( \sqrt{d+1} / d! \, 2^{d/2} ) \, l^d \equiv c_d l^d $.
Doing the single surviving integral over $l^2$, 
$\int_0^\infty dl^2 \, \exp (- \lambda c_d l^d ) = 
(c_d \lambda)^{-2/d} \Gamma ((d+2)/d)$, gives
$< \! V \! > = 2 / d \lambda = c_d \, l_0^d $.
Solving this last expression for $l_0^2$ then gives the desired result
\beq
l_0^2 = { 1 \over \lambda^{2/d} } \left [ 
{2 \over d } \, { d! \, 2^{d/2} \over \sqrt{d+1} } \right ]^{2/d}
\label{eq:l2d}
\eeq
(which, for example, gives $l_0 = 2.153$ for $\lambda=1$ in four dimensions,
in reasonable agreement with the actual value 
$l_0 \approx 2.43$ found in \cite{critical} near the transition point).
The result of Eq.~(\ref{eq:kc4}), extended to $d$ dimensions, should then read
\beq
k_c \; = \; { \lambda \, l_0^d \over \, d^3 \, l_0^{d-2} } \; = \;
{ \lambda \, l_0^2 \over \, d^3 } \;\; ,
\label{eq:kcd}
\eeq
which is in fact the same result as before in $d=4$.
Using Eq.~(\ref{eq:l2d}) inserted into Eq.~(\ref{eq:kcd})
one then obtains in the large $d$ limit for the naturally dimensionless
combination $k / \lambda^{(d-2)/d}$ 
\beq
{ k_c \over \lambda^{d-2 \over d} } \; = \; 
{ 2^{1+ {2 \over d} } \over d^3 } \, \left [ 
{ \Gamma (d) \over \sqrt{d+1} } \right ]^{2/d} \sim {2 \over e^2 } \,
{1 \over d } \;\; .
\label{eq:kcd1}
\eeq
This result would then lead us to conclude that the above 
critical dimensionless ratio
of couplings is given in the large-$d$ limit by 
$k_c/\lambda \sim 1 / d $.
One should be careful though not to assign any deep physical significance
to this result, which is only meant to help determine the critical
values for the bare coupling constants.

In the following we will now revert back, for simplicity, to the case of
an expansion about $l_0=1$.
Returning to the partition function (and averages derived from it)
associated with Eq.~(\ref{eq:beta1}), 
we note that it can be formally computed via
\beq
Z \; = \; \int \, \prod_{i=1}^N \, d \e_i \, 
e^{ - \e \, M \, \e }
\; = \; { \pi^{N/2} \over \sqrt{ \det M } }
\; = \; { \pi^{N/2} \over \sqrt{ \prod_{i=1}^N \, \lambda_i } } \;\; ,
\eeq
with $N=2d(d+1)$.
Convergence of the Gaussian integral then requires $k d^3 > \lambda$.
From 
\beq
\ln Z \; = \; { N \over 2 } \ln \pi \, - \,
{ 1 \over 2 } \, \sum_{i=1}^N \, \ln \lambda_i 
\; \sim \; 
{ N \over 2 } \ln \pi \, - \,
{ 1 \over 2 } \, \int_0^\infty \, ds \, \rho (s) \, \ln \lambda (s)
\;\; ,
\eeq
and using the fact that for the cross polytope to leading order in $1/d$
all eigenvalues are equal, one has
\beq
\log Z \; = \; { \sqrt{d} \, 2^{{d \over 2}+1} \over d! } \, 
\left ( k \, d^3 \, - \, \lambda \right ) \, + \,
d \, (d+1) \, \log \left [ 
8 \pi  /  { \sqrt{d} \, 2^{{d \over 2} +1} \over d! } \,
\left ( k \, d^3 \, - \, \lambda \right )  \right ] \;\; ,
\eeq
with the first term arising from the constant term in the action,
and the second term from the $\e$-field Gaussian integral.
Therefore the general structure, to leading order in the weak
field expansion at large $d$, is 
$ \log Z  = c_1 ( k \, d^3 - \lambda ) - 
d(d+1) \log ( k d^3 - \lambda ) + c_2 $
with $c_1$ and $c_2$ $d$-dependent constants,
and therefore 
$\partial^2 \log Z / \partial k^2 \sim 1 / ( k d^3 - \lambda )^2 $
with divergent curvature fluctuations in the vicinity of the critical point 
at $k d^3 = \lambda$.

\subsection{Inclusion of higher order terms}

It seems legitimate to ask what happens if the fluctuations
in the $\e$'s are large enough so that the quadratic
approximation is no longer adequate.
Then one has from  Eq.~(\ref{eq:beta1}), to lowest order in $1/d$, 
\beq
{ \sqrt{d} \, 2^{d/2} \over d! } \, 2 \, \left [
\left ( \lambda \, - \, k \, d^3 \right ) \,  
\left ( 1 \, - \, {1 \over 8} \, \sum \e_{ij}^2 \, + \, \dots \right )
\, + \, {\sigma \, \lambda \over 16} \, \sum \e_{ij}^4 \,
\right ] \;\; ,
\label{eq:beta3}
\eeq
where we have again included a cutoff term, proportional to $\sigma$, for   
each edge.
Then, again up to the constant term and an overall multiplicative
factor, the action reduces to
\beq
- \, { 1 \over 2 } \, \left ( \lambda \, - \, k \, d^3 \right ) \, 
\sum \e_{ij}^2 \, + \,
{\sigma \, \lambda \over 4} \, \sum \e_{ij}^4 \;\; .
\label{eq:beta4}
\eeq
At strong coupling $k \rightarrow 0$, the minimum lies at
a non-vanishing value of the $\e$'s, 
namely $\e_{ij} = \pm 1 / \sqrt{\sigma}$.
Since we started out with equilateral simplices with unit edges,
this result is telling us that the edges have to be slightly extended (or
shortened) to reach the minimum. 
As $k$ is increased, the minimum eventually moves to
the origin for $k = \lambda / d^3 $.
Neglecting the effects of fluctuations in
the $\e$ fields, $< \e \cdot \e > - < \e >^2 =0 $, which is similar
to the Landau treatment of ferromagnetic transitions, 
one then obtains
\beq
- \, {1 \over 2} \, \left ( \lambda \, - \, k \, d^3 \, \right ) \, \e^2 
\, + \, { \sigma \, \lambda \over 4 } \, \e^4 \;\; .
\label{eq:landau}
\eeq
For $k d^3 > \lambda $ the minimum is at the origin, whereas
for $k d^3 < \lambda $ it moves away from it.
For $\lambda > k d^3 $ one has a shifted minimum at 
$\e_0 \, = \, \pm (1 - k d^3 / \lambda \sigma )^{1/2}$
and a total action 
$I(\e_0) \, = \, - \, \lambda \, (1 - k d^3 / \lambda )^2 / 4 \sigma $.
As a result $\e_0$ vanishes at $k= \lambda / d^3 $,
and so does $I(\e_0) $.

If we apply the ideas of mean field theory, we need to keep the terms of 
order $1/d$ in Eq.~(\ref{eq:beta}). In the $\epsilon_{ij}\epsilon_{ik}$
term, we assume that the fluctuations are small and replace $\epsilon_{ik}$ 
by its average $\bar{\e}$.
Each $\epsilon_{ij}$ has $4d-2$ neighbors (edges with
one vertex in common with it); this has to be divided by $2$ to avoid double 
counting in the sum, so the contribution is $(2d-1) \, \bar{\e}$.
Then to lowest order in $1/d$, the action is proportional to
\beq
\left ( \lambda \, - \, k \, d^3 \right ) \,
\left [ 1 \, - \, {1 \over 8} \, \sum \e_{ij}^2 \, + \,
{ 1 \over 4 } \, \bar{\e} \sum \e_{ij} \right ] \;\; .
\eeq
This gives rise to the same partition function as obtained earlier, and 
using it to calculate the average value of $\epsilon_{ij}$ gives $\bar{\e}$,
as required for consistency.

To summarize, in this section we have developed an expansion
in power of $1/d$, which relies on a combined and simultaneous 
use of the weak field expansion.
It can therefore be regarded as a double expansion in $1/d$
and $\e$, valid wherever the fields are smooth enough and the
geometry is close to flat, which presumably is the
case to some extent at large distances in the vicinity of the
lattice critical point at $k_c$.
In the next section we will develop a different and complementary
$1/d$ expansion, which will not require weak fields, but will rely
instead on the strong coupling (small $k=1/8 \pi G$, or large $G$) limit.
As such it should now be considered as a double expansion in $1/d$ and $k$.
Its validity will be in a regime where the fields are not smooth, and
in fact will rely on considering lattice gravitational field configurations
which are very far from smooth at short distances.

\vskip 30pt
\newsection{Strong Coupling Expansion in Large Dimensions}
\hspace*{\parindent}

In this section we discuss the strong coupling (small $k=1/(8 \pi G)$)
expansion of the lattice gravitational partition function, first
in the general case, and subsequently for large $d$.
The resulting series is expected to be useful up to some $k=k_c$,
where $k_c$ is the lattice critical point 
(as determined for example from Eq.~(\ref{eq:kcd1})),
at which the partition function develops a singularity.
It appears that the phase $ k > k_c$ is of limited physical interest, since
in that phase spacetime collapses into a two-dimensional
manifold \cite{hw84,lesh,monte} (in fact, one of the first examples
of compactification due to non-perturbative dynamics, as opposed
to a specific choice of boundary conditions).

There will be two main aspects to the following discussion.
The first aspect will be the development of a systematic expansion
for the partition function and the correlation functions
in powers of $k$, and a number of rather general considerations that follow
from it.
The second main aspect will be a detailed analysis and
interpretation of the individual terms which appear order by order 
in the strong coupling expansion. 
This second part will then lead to a discussion of what
happens for large $d$.

\subsection{The measure}

We will therefore first focus on the four-dimensional case, and
then later exhibit its more or less immediate generalization to $d>4$.
The 4-dimensional Euclidean lattice action \cite{regge,hw84,hartle} contains
the usual cosmological constant and Regge scalar curvature terms
\beq 
I_{latt} \; = \;  \lambda \, \sum_h V_h (l^2) \, - \, 
k \sum_h \delta_h (l^2 ) \, A_h (l^2) \;\; , 
\label{eq:ilatt} 
\eeq
with $k=1/(8 \pi G)$, and possibly additional higher derivative terms
as well.
The action only couples edges which belong either to
the same simplex or to a set of neighboring simplices, and can therefore
be considered as {\it local}, just like the continuum action. 
It leads to a lattice partition function defined as
\beq 
Z_{latt} \; = \;  \int [ d \, l^2 ] \; e^{ 
- \lambda \sum_h V_h \, + \, k \sum_h \delta_h A_h } \;\; ,
\label{eq:zlatt} 
\eeq
where, as customary, the lattice ultraviolet cutoff is set equal to one
(i.e. all length scales are measured in units of the lattice cutoff).
For definiteness the measure will be of the form \cite{hw84,hartle,cms1}
\beq
\int [ d \, l^2 ] \; = \;
\int_0^\infty \; \prod_s \; \left ( V_d (s) \right )^{\sigma} \;
\prod_{ ij } \, dl_{ij}^2 \; \Theta [l_{ij}^2] \;\; .
\eeq
The lattice partition function $Z_{latt}$ should be compared to the
continuum Euclidean Feynman path integral
\beq
Z_{cont} \; = \; \int [ d \, g_{\mu\nu} ] \; e^{ 
- \lambda \, \int d x \, \sqrt g \, + \, 
{ 1 \over 16 \pi G } \int d x \sqrt g \, R} \;\; ,
\label{eq:zcont}
\eeq
which involves a functional integration
over all metrics, with functional measure ~\cite{dewittlec,dewitt,misner}
\beq
\int [ d \, g_{\mu\nu} ] \; = \; \int \prod_x \; 
\left ( g(x) \right )^{ (d-4)(d+1)/8 } \;
\prod_{\mu \ge \nu} \, d g_{\mu \nu} (x)
\; \mathrel{\mathop\rightarrow_{ d = 4}} \;
\int \prod_x \prod_{\mu \geq \nu} d g_{\mu \nu} (x) \;\; . 
\label{eq:dewitt}
\eeq
Since we will be doing an expansion in the kinetic term
proportional to $k$, it will be convenient to include the
$\lambda$-term in the measure. 
We will set therefore in this Section
\beq 
d \mu (l^2) \; \equiv \; [ d \, l^2 ] \, e^{- \lambda \sum_h V_h }
\;\; .
\label{eq:mulatt} 
\eeq
It should be clear that this last expression represents a fairly non-trivial 
quantity, both in view of the relative complexity
of the expression for the volume of a simplex, Eq.~(\ref{eq:vol}),
and because of the generalized triangle inequality constraints 
already implicit in $[d\,l^2]$.
But, like the continuum functional measure, it is certainly {\it local},
to the extent that each edge length
appears only in the expression for the volume of those simplices
which explicitly contain it.
Also, we note that in general the integral $\int d \mu$ can only be
evaluated numerically; nevertheless this can be done, at least in principle, 
to arbitrary precision.
Furthermore, $\lambda$ sets the overall scale and can therefore be set 
equal to one without any loss of generality
(one can also conveniently normalize the integration measure, so that 
$Z_0 \equiv \int d \mu (l^2) =1$, but this will not be necessary here).

To summarize, the effective strong coupling measure of 
Eq.~(\ref{eq:mulatt}) has the properties
that 1) it is local in the lattice metric of Eq.~(\ref{eq:latmet}),
to the same extent that the continuum measure is ultra-local, 
2) it restricts all edge lengths to be positive, and 3) it imposes
a soft cutoff on large simplices due to the $\lambda$-term and the
generalized triangle inequalities.
Apart from these constraints, it 
does {\it not} significantly restrict the fluctuations
in the lattice metric field at short distances.
It will be the effect of the curvature term to restrict such
fluctuation, by coupling the metric field between simplices, in
the same way as the derivatives appearing in the continuum Einstein
term couple the metric between infinitesimally close spacetime points.

\subsection{Expansion in powers of $k$}

From now on we will discuss $Z_{latt}$ only, and drop the subscript
$latt$.
As a next step, $Z$ is expanded in powers of $k$,
\beq 
Z (k) \; = \;  \int d \mu (l^2) \, \; e^{k \sum_h \delta_h \, A_h } 
\; = \;  \sum_{n=0}^{\infty} \, { 1 \over n!} \, k^n \, 
\int d \mu (l^2) \, \left ( \sum_h \delta_h \, A_h \right )^n \;\; .
\eeq
It is easy to show that $Z (k) \, = \, \sum_{n=0}^{\infty} a_n \, k^n $
is analytic at $k=0$, so this expansion is well defined, up to the
nearest singularity in the complex $k$ plane.
An estimate for the expected location of such a singularity in the large-$d$
limit was given in Eq.~(\ref{eq:kcd1}) of the previous section.
Beyond this singularity $Z(k)$ can sometimes be extended, for example, 
via Pad\'e or differential approximants \cite{dombgreen,pade}
\footnote{
It is well known that a first order transition cannot affect
the singularity structure of $Z(k)$ as viewed from the strong coupling phase,
as the free energy is $C_\infty$ at a first order transition.
$Z(k)$, as defined from the strong coupling phase, will be
non-analytic only at the second order, endpoint transition, 
modulo an exponentially small imaginary part appearing in the
metastable phase, if one exists.
Approaching the phase transition from the strong coupling side
detects the physically relevant end-point singularity, where the
correlation length diverges and scale invariance is
presumably recovered \cite{critical}. 
}.
The above expansion is of course analogous to the high temperature expansion
in statistical mechanics systems,
where the on-site terms are treated exactly and
the kinetic or hopping term is treated as a perturbation.
Singularities in the free energy or its derivatives can usually
be pinned down with the knowledge of a large enough number of terms in the
relevant expansion \cite{dombgreen}.
The often surprisingly rich structure of singularities
in the complex coupling plane and their volume dependence
has been explored in detail for some simple exactly soluble
models with a finite number of degrees of freedom \cite{pearson}.

Next consider a fixed, arbitrary hinge on the lattice, and
call the corresponding curvature term in the action $\delta A $.
Such a contribution will be denoted in the following, as is customary
in lattice gauge theories, a {\it plaquette} contribution.
For the average curvature on that hinge one has
\beq 
< \delta A > \; = \; 
{\displaystyle \sum_{n=0}^{\infty} \, { 1 \over n!} \, k^n \, 
\int d \mu (l^2) \, \delta A \, \left ( \sum_h \delta_h \, A_h \right )^n
\over
\displaystyle \sum_{n=0}^{\infty} \, { 1 \over n!} \, k^n \, 
\int d \mu (l^2) \, \left ( \sum_h \delta_h \, A_h \right )^n } \;\; .
\label{eq:rseries}
\eeq
After expanding out in $k$ the resulting expression, one
obtains for the cumulants
\beq
< \delta \, A > \; = \; \sum_{n=0}^{\infty} c_n \, k^n \;\; ,
\label{eq:rseries2}
\eeq
with 
\beq
c_0 \; = \; 
{\displaystyle 
\int d \mu (l^2) \, \delta \, A 
\over
\displaystyle \int d \mu (l^2) } \;\; ,
\label{eq:k0}
\eeq
whereas to first order in $k$ one has
\beq
c_1 \; = \; 
{\displaystyle 
\int d \mu (l^2) \, \delta \, A \, \left ( \sum_{h} \delta_{h} \, A_{h} \right )
\over
\displaystyle \int d \mu (l^2) }
\; - \;
{\displaystyle 
\int d \mu (l^2) \, \delta \, A \, \cdot 
\int d \mu (l^2) \, \sum_{h} \delta_{h} \, A_{h}
\over
\displaystyle \left ( \int d \mu (l^2) \right )^2 } \;\; .
\label{eq:k1}
\eeq
This last expression clearly represents a measure of the fluctuation
in $\delta \, A$, namely 
$[ \langle ( \sum_{h} \delta_{h} \, A_{h} )^2 \rangle - 
\langle \sum_{h} \delta_{h} \, A_{h} \rangle^2 ] / N_{h} $, 
using the homogeneity properties of the lattice
$\delta A \rightarrow \sum_{h} \delta_{h} A_{h} / N_{h} $.
Equivalently, it can be written in an even more compact way as
$N_{h} [ \langle (\delta A)^2 \rangle - \langle \delta A \rangle^2 ] $. 
To second order in $k$ one has
\bea
c_2 \; = \; && 
{\displaystyle 
\int d \mu (l^2) \, \delta A \, \left ( \sum_{h} \delta_{h} \, A_{h} \right )^2
\over
2 \, \displaystyle \int d \mu (l^2) }
\; - \;
{\displaystyle 
\int d \mu (l^2) \, \sum_{h} \delta_{h} \, A_{h} \cdot
\int d \mu (l^2) \, \delta A \, \sum_{h} \delta_{h} \, A_{h}
\over
\displaystyle \left ( \int d \mu (l^2) \right )^2 }
\nonumber \\
\; & - & \;
{\displaystyle 
\int d \mu (l^2) \, \left ( \sum_{h} \delta_{h} \, A_{h} \right )^2
\int d \mu (l^2) \, \delta A
\over
2 \, \displaystyle \left ( \int d \mu (l^2) \right )^2 }
\; + \;
{\displaystyle 
\int d \mu (l^2) \, \delta A \, \cdot \left (
\int d \mu (l^2) \, \sum_{h} \delta_{h} \, A_{h} \right )^2
\over
\displaystyle \left ( \int d \mu (l^2) \right )^3 } 
\label{eq:k2}
\eea
which now corresponds to
$c_2 = N_{h}^2 \, [ \langle (\delta A)^3 \rangle - \, 3 \langle \delta A \rangle \langle ( \delta A )^2 \rangle
+ \,  2 \langle \delta A \rangle^3 ]/2 $.
At the next order one has 
$c_3 = N_{h}^3 \, [ \langle (\delta A)^4 \rangle - \, 4 \langle \delta A \rangle \langle ( \delta A )^3 \rangle
- \, 3 \langle (\delta A)^2 \rangle^2 
+ \, 12 \langle (\delta A)^2 \rangle \langle \delta A \rangle^2
- \,  6 \langle \delta A \rangle^4 ]/6 $, 
and so on.
Note that the expressions in square parentheses become rapidly quite small,
$O(1/N_h^n)$ with increasing order $n$, as a result of large 
cancellations that must arise eventually between individual terms
inside the square parentheses.
In principle, a careful and systematic numerical evaluation of the above
integrals (which is quite feasible in practice) would allow the
determination of the expansion coefficients in $k$ for the average
curvature $<\delta A>$ to rather high order, but we shall not pursue
this line of inquiry here
\footnote{
As an example, consider a non-analyticity in the average scalar curvature
\beq
{\cal R} (k) \; = \; 
{ < \int d x \, {\textstyle {\sqrt{g(x)}} \displaystyle} \, R(x) >
\over < \int d x \, {\textstyle {\sqrt{g(x)}} \displaystyle} > } \;\; ,
\label{eq:rave}
\eeq
assumed to be of the form of an algebraic singularity at $k_c$, namely
$ {\cal R} (k) \; \mathrel{\mathop\sim_{ k \rightarrow k_c}} \;
A_{\cal R} \, ( k_c - k )^\delta $.
It will lead to a behavior, for the general term in the series in $k$,
of the type
\beq
(-1)^n \, A_{\cal R} \, { (\delta -n +1) (\delta -n +2) \dots \delta 
\over n! \, k_c^{n - \delta} } \, k^n \;\; .
\eeq
Given enough terms in the series, the singularity structure can
then be investigated using a variety of increasingly sophisticated
methods \cite{dombgreen,qftexp,guida,campo}.
In Ref. \cite{critical} the curvature ${\cal R}(k)$ was
computed numerically for various values of $k$, from which one
can extract an approximate value for the
coefficients, namely 
${\cal R}(k)$=$- 9.954 + 62.11 k + 195.94 k^2 - 1340.65 k^3 + 40483.75 k^4
+O(k^5)$.
A better and much more accurate way would be a direct determination of
the individual coefficients, via the edge length integrals
of Eqs.~(\ref{eq:k1}) and (\ref{eq:k2}).
}.

It is advantageous to isolate in the above expressions the {\it local}
fluctuation term, from those terms that involve {\it correlations} between
different hinges.
To see this, one needs to go back, for example,
to the first order expression in Eq.~(\ref{eq:k1})
and isolate in the sum $\sum_{h}$ the contribution which
contains the selected hinge with value $\delta A$, namely
\beq
\sum_{h} \delta_{h} \, A_{h} \; = \;
\delta \, A \, + \, \sum_{h}\, ' \delta_{h} \, A_{h} \;\; ,
\eeq
where the primed sum indicates that the term containing $\delta A$
is {\it not} included.
The result is
\bea
c_1 \; = \; && {\displaystyle 
\int d \mu (l^2) \, ( \delta \, A )^2 
\over
\displaystyle \int d \mu (l^2) }
\; - \;
{\displaystyle 
\left ( \int d \mu (l^2) \, \delta \, A \, \right )^2 
\over
\displaystyle \left ( \int d \mu (l^2) \right )^2 }
\nonumber \\
&& \; + \;
{\displaystyle 
\int d \mu (l^2) \, \delta \, A \, 
\sum_{h}\, ' \delta_{h} \, A_{h}
\over
\displaystyle \int d \mu (l^2) }
\; - \;
{\displaystyle 
\left ( \int d \mu (l^2) \, \delta \, A \right ) \, \left (
\int d \mu (l^2) \, \sum_{h}\, ' \delta_{h} \, A_{h} \right )
\over
\displaystyle \left ( \int d \mu (l^2) \right )^2 } \;\; .
\label{eq:k1ex}
\eea
One then observes the following: the first two terms describe the
{\it local} fluctuation of $\delta A$ on a given hinge;
the third and fourth terms describe {\it correlations} between $\delta A$
terms on {\it different} hinges.
But because the action is {\it local}, the only non-vanishing
contribution to the last two terms comes from edges and hinges 
which are in the immediate vicinity of the hinge in question.
For hinges located further apart (indicated below by ``$not \, nn$'')
one has that their fluctuations
remain uncorrelated, leading to a vanishing variance 
\beq
{\displaystyle 
\int d \mu (l^2) \, \delta \, A \, 
\sum_{h \, not \, nn}\, ' \delta_{h} \, A_{h}
\over
\displaystyle \int d \mu (l^2) }
\; - \;
{\displaystyle 
\left ( \int d \mu (l^2) \, \delta \, A \right ) \, \left (
\int d \mu (l^2) \, \sum_{h \, not \, nn}\, ' \delta_{h} \, A_{h} \right )
\over
\displaystyle \left ( \int d \mu (l^2) \right )^2 } \; = \; 0 \;\; ,
\label{eq:variance}
\eeq
since for uncorrelated random variables $X_n$'s, $<X_n X_m>-<X_n><X_m>=0$.
Therefore the only non-vanishing contributions in the last two terms
in Eq.~(\ref{eq:k1ex}) come from hinges which are {\it close} to each other.

The above discussion makes it clear that a key quantity is the
{\it correlation} between different plaquettes,
\beq
< ( \delta \, A )_{h} \, ( \delta \, A)_{h'} > \; = \; 
{\displaystyle 
\int d \mu (l^2) \,
 ( \delta \, A)_{h} \, ( \delta \, A)_{h'} \,
e^{k \sum_h \delta_h \, A_h }
\over
\displaystyle \int d \mu (l^2) \, e^{k \sum_h \delta_h \, A_h } } \;\;
,
\label{eq:corr}
\eeq
or, better, its {\it connected} part (denoted here by $< \dots >_C$)
\beq
< ( \delta \, A )_{h} \, ( \delta \, A)_{h'} >_C \;\; \equiv \;\;
< ( \delta \, A )_{h} \, ( \delta \, A)_{h'} > \; - \;
< ( \delta \, A )_{h} > \, < ( \delta \, A)_{h'} > \;\; ,
\label{eq:corrconn}
\eeq
which subtracts out the trivial part of the correlation.
Here again the exponentials in the numerator and denominator can be
expanded out in powers of $k$, as in Eq.~(\ref{eq:rseries}).
The lowest order term in $k$ will involve the correlation
\beq
\int d \mu (l^2) \, ( \delta \, A)_{h} \, ( \delta \, A)_{h'} \;\; .
\eeq
But unless the two hinges are close to each other, they will fluctuate
in an uncorrelated manner, with
$< ( \delta \, A )_{h} \, ( \delta \, A)_{h'} > -
< ( \delta \, A )_{h} > < ( \delta \, A)_{h'} > \, = \, 0 $.
In order to achieve a non-trivial correlation, the path between
the two hinges 
$h$ and $h'$ needs to be tiled by at least as many terms from
the product $ ( \sum_h \delta_h \, A_h )^n $ in
\beq
\int d \mu (l^2) \, ( \delta \, A)_{h} \, ( \delta \, A)_{h'} \,
\left ( \sum_h \delta_h \, A_h \right )^n
\eeq
as are needed to cover the distance $l$ between the two hinges.
One then has
\beq
< ( \delta \, A )_{h} \, ( \delta \, A)_{h'} >_C \; \sim \; 
k^l \; \sim \; e^{- l / \xi } \;\; ,
\label{eq:kxi}
\eeq
with the correlation length $ \xi = 1 / | \log k | \rightarrow 0 $
to lowest order as $k \rightarrow 0 $
(here we have used the usual definition of the correlation length $\xi$,
namely that a generic correlation function
is expected to decay as $ \exp (- {\rm distance} / \xi) $ for
large separations)
\footnote{This statement, taken literally, oversimplifies the situation a bit, 
as depending on the spin (or tensor structure) of the operator appearing in the
correlation function, the large distance decay of the corresponding
correlator is determined by the lightest excitation in that specific channel.
But in the gravitational context one is mostly concerned with correlators
involving spin two (transverse-traceless) objects.
}.
This last result is quite general, and holds for example irrespective
of the boundary conditions (unless of course $\xi \sim L$, where $L$ is the
linear size of the system, in which case a path can be found
which wraps around the lattice).

But further thought reveals that the above result is in fact not
completely correct, due to the fact that in order to achieve
a non-vanishing correlation one needs, at least to lowest order,
to connect the two hinges by a narrow tube. 
The previous result should then read correctly as 
\beq
< ( \delta \, A )_{h} \, ( \delta \, A)_{h'} >_C \; \sim \; 
\left ( k^{n_d} \right )^l \;\; ,
\label{eq:nd}
\eeq
where, as will be shown in more detail below,
$n_d \, l$ represents (approximately) the minimal number of dual
lattice polygons that are needed to form a closed surface connecting
the hinges $h$ and $h'$, with $l$ the actual distance (in lattice units) 
between the two hinges.

\subsection{Rotation matrices, Voronoi loops and closed surfaces}

Up to this point our considerations have been quite general, and
therefore do not take into account yet the detailed nature of
the local interaction expressed in the action term $\sum_h \delta_h A_h $.
It is well known that the deficit angle $\delta_h$ describes
the rotation of a vector $V^{\mu}$ parallel transported
around a closed loop encircling the hinge $h$.
This full rotation is best described in terms of a (Lorentz) rotation
matrix $\bf R$, an element of $SO(4)$ or $SO(3,1)$, depending on the
signature of the metric, and whose matrix elements will depend
on the specific choice of coordinates at the point in question.
In $d$ dimensions the corresponding objects would be $SO(d)$ or $SO(d-1,1)$
rotations, in the Riemannian and pseudo-Riemannian case respectively
\footnote{
The preceding observations can in fact be developed into a consistent
first order (Palatini) formulation of Regge gravity, with suitably chosen
independent transformation matrices and metrics, related to each other
by a set of appropriate lattice equations of motion \cite{caselle}. 
One would expect the first and second order formulations to
ultimately describe the same quantum theory, with common universal
long-distance properties.
How to consistently define finite rotations, frames
and connections in Regge gravity was first discussed
systematically in \cite{froh}.
}.

Just as in the continuum, where the affine connection and therefore
the infinitesimal rotation matrix is determined by the metric
and its first derivatives, on the lattice the elementary
rotation matrix between simplices ${\bf R}_{s,s+1}$ is fixed by the
difference between the $g_{ij}$'s of Eq.~(\ref{eq:latmet}) within
neighboring simplices.
Consider therefore a closed path 
$\Gamma$ encircling a hinge $h$ and passing through
each of the simplices that meet at that hinge.
In particular one may take $\Gamma$ to be the boundary of the polyhedral
dual (or Voronoi) area surrounding the hinge \cite{hw84}.
We recall that the Voronoi polyhedron dual to a vertex $P$ is the set
of all points on the lattice which are closer to $P$ than any other vertex;
the corresponding new vertices then represent the sites on the dual lattice.
A unique closed parallel transport path can then be assigned to each hinge,
by suitably connecting sites in the dual lattice.

With each neighboring pair of simplices $s,s+1$ one associates a
Lorentz transformation ${\bf R}^{\alpha}_{\;\; \beta}$, which describes
how a given vector $ V_\mu $ transforms between the local coordinate
systems in these two simplices,
\beq
{V '}^\alpha \; = \; 
\Bigl [ {\bf R}_{s,s+1} \Bigr ]^{\alpha}_{\;\; \beta} \, V^\beta \;\; .
\label{eq:rot}
\eeq
The above Lorentz transformation is then directly related to the continuum
path-ordered ($P$) exponential of the integral of the local affine connection
$ \Gamma^{\lambda}_{\mu \nu}$ via
\beq
{\bf R}^\alpha_{\;\; \beta} \; = \; \Bigl [ P e^{\int_
{{\bf path \atop between \; simplices}}
\Gamma_\lambda d x^\lambda} \Bigr ]^\alpha_{\;\; \beta}  \;\; .
\label{eq:rot1}
\eeq
Next consider moving a vector $V$ once around a Voronoi loop,
i.e. a loop formed by Voronoi edges surrounding a chosen hinge.
The change in the vector $V$ is given by
\beq
\delta V^\alpha = ({\mathbf{R-1}})^{\alpha}_{\;\;\beta}V^\beta \;\; ,
\eeq
where ${\bf R} \equiv \prod_s {\bf R}_{s,s+1} $
is now the total rotation matrix associated with the given hinge. 
Since in the continuum $\delta V$ is given by
$\delta V^\alpha = \half \; R^{\alpha}_{\;\;\beta\mu\nu} \; 
A^{\mu\nu}_{\Gamma} \; V^\beta $ ,
where $A^{\mu\nu}_{\Gamma}$ is the antisymmetric bivector
representing the loop area, one has the identification
\beq
\half \; R^{\alpha}_{\;\;\beta\mu\nu}A^{\mu\nu}_{\Gamma} \; = \;
({\mathbf{R-1}})^{\alpha}_{\;\;\beta} \;\; .
\label{eq:riemrot}
\eeq
To first order in the deficit angle $\delta$ one then recovers
the well known result
\beq
R^{\alpha}_{\;\;\beta\mu\nu} \; = \;
\frac{\delta}{A_\Gamma } \; U^{\alpha}_{\;\;\beta} \, U_{\mu\nu} \;\; ,
\eeq
where $U_{\alpha\beta}$ represents the hinge bivector,
$ U_{\alpha\beta}= { 1 \over 2 A_h } \,
\e_{\alpha\beta\mu\nu} \, l_1^{\mu}l_2^{\nu} $,
with $l_1$ and $l_2$ the two hinge vectors and $A_h$ the area of the hinge,
and use has been made of the relationship between the original volumes
and their dual counterparts,
$ A_{\alpha\beta}^{\Gamma} U^{\alpha\beta} = 2 A_{\Gamma} $.
As a result, one can relate the deficit angle directly to
the effect of a complete rotation of a vector around a hinge,
\beq
\Bigl [ \prod_s {\bf R}_{s,s+1}   \Bigr ]^{\mu}_{\;\; \nu} \; = \;
\Bigl [ \, e^{\delta_h U_{..}^{(h)}} \Bigr ]^{\mu}_{\;\; \nu}  \;\; .
\label{eq:fullrot}
\eeq
In other words, the product of rotation matrices around the closed
elementary loop describes a rotation in a plane perpendicular to
the hinge, by an angle $\delta_h$.
Equivalently, this last expression can be re-writtten in terms of a surface
integral of the Riemann tensor, projected along the surface area element
bivector $A^{\alpha\beta}_{\Gamma}$ associated with the loop,
\beq
\Bigl [ \prod_s {\bf R}_{s,s+1}   \Bigr ]^{\mu}_{\;\; \nu} \; \approx \;
\Bigl [ \, e^{\half \int_S 
R^{\, \cdot}_{\;\; \cdot \, \alpha\beta} \, A^{\alpha\beta}_{\Gamma} } 
\Bigr ]^{\mu}_{\;\; \nu}  \;\; .
\label{eq:rotriem}
\eeq

\begin{center}
\epsfysize=8.0cm
\epsfbox{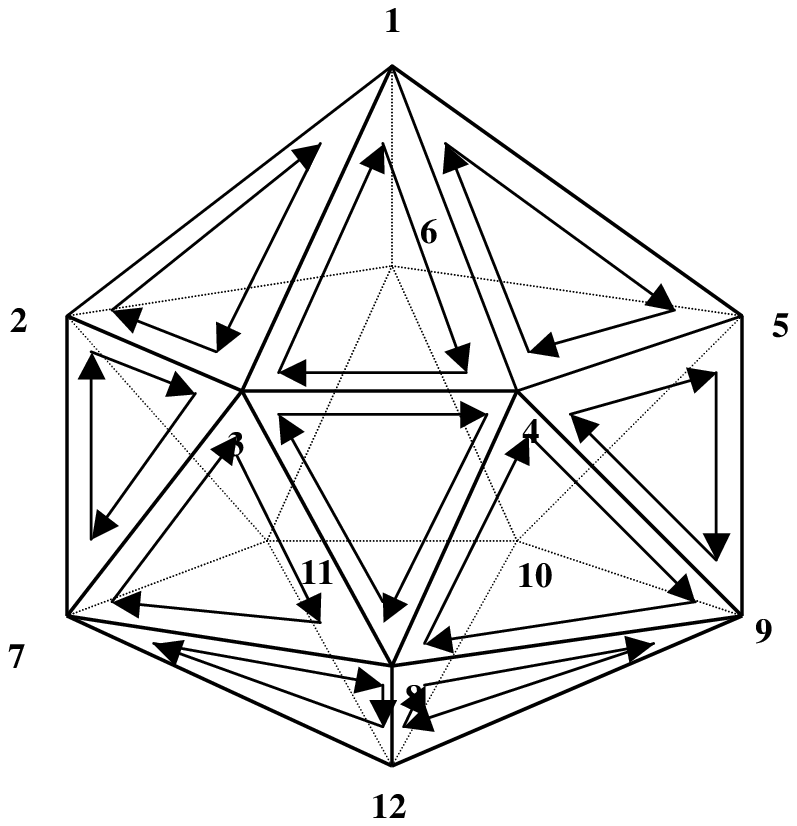}
\end{center}

\noindent{\small Fig 2. Elementary closed surface tiled with parallel
transport polygons.
For each link of the dual lattice, the parallel transport matrices ${\bf R}$
are represented by an arrow. 
In spite of the fact that the Lorentz matrices ${\bf R}$ fluctuate with
the local geometry, two contiguous, oppositely oriented arrows always
give ${\bf R} {\bf R}^{-1} = 1$. \medskip }

Let us now return to the strong coupling expansion, and it will
be advantageous now to focus on general properties of the 
parallel transport matrices ${\bf R}$
\footnote{
The role of continuous rotation matrices in Regge gravity is brought out
in a particularly clear way by the first order approach of Ref. \cite{caselle}.
}.
For smooth enough geometries, with small curvatures, the above 
rotation matrices can be chosen to be close to the identity.
Small fluctuations in the geometry will then imply small deviations
in the ${\bf R}$'s from the identity matrix.
But for strong coupling ($k \rightarrow 0$) it
was already emphasized before that the
measure $\int d \mu (l^2)$ does not significantly restrict fluctuations
in the lattice metric field.
As a result these fields can be regarded in this regime 
as basically unconstrained random variables, only subject
to the relatively mild constraints implicit in the measure $d \mu$.
The geometry is generally far from smooth
since there is no coupling term to enforce long range
order (the coefficient of the lattice Einstein term is zero),
and one has as a consequence large local fluctuations in the geometry.
The matrices ${\bf R}$ will therefore fluctuate with the local
geometry, and average out to zero, or a value close to
zero
\footnote{ 
In the sense that, for example, the $SO(4)$ rotation
$ {\bf R}_\theta = \left ( \matrix{ \cos \theta & - \sin \theta & 0 & 0 \cr
                   \sin \theta &   \cos \theta & 0 & 0 \cr 
                   0 & 0 & 1 & 0 \cr     0 & 0 & 0 & 1 \cr 
} \right ) $ 
averages out to zero when integrated over $\theta$. 
In general an element of $SO(n)$ is described by $n(n-1)/2$ independent 
parameters, which in the case at hand can be conveniently chosen as the six
$SO(4)$ Euler angles.
The uniform (Haar) measure over the group is then
$d \mu_H ( {\bf R} ) = {1 \over 32 \pi^9 }
\int_0^{2 \pi} d \theta_1 \int_0^{\pi} d \theta_2 \int_0^{\pi} d \theta_3
\int_0^{\pi} d \theta_4 \sin \theta_4 \int_0^{\pi} d \theta_5 \sin \theta_5
\int_0^{\pi} d \theta_6 \sin^2 \theta_6$.
This is just a special case of the general $n$ result \cite{haar}, which reads
$d \mu_H ( {\bf R} ) = 
\left ( \prod_{i=1}^n \Gamma (i/2) / 2^n \, \pi^{n(n+1)/2} \right )
\prod_{i=1}^{n-1} \prod_{j=1}^i 
\sin^{j-1} \theta_{\; j}^i \, d \theta_{\; j}^i $
with $ 0 \le \theta_{\; k}^1 < 2 \pi $, $ 0 \le \theta_{\; k}^j < \pi $.
}.

This is quite similar of course to what happens in $SU(N)$ Yang-Mills theories,
or even more simply in (compact) QED, where the analogs of the $SO(d)$ rotation
matrices ${\bf R}$ are phase factors $U_{\mu}(x)=e^{iaA_{\mu}(x)}$.
One has there $\int { d A_{\mu} \over 2 \, \pi } \, U_{\mu}(x) = 0 $
and 
$\int { d A_{\mu} \over 2 \, \pi } \, U_{\mu}(x) \, U^{\dagger}_{\mu}(x) = 1 $.
In addition, for two contiguous closed paths $C_1$ and $C_2$
sharing a common side one has
\beq
e^{ i \oint_{C_1} {\bf A \cdot dl } } \, e^{ i \oint_{C_2} {\bf A \cdot dl } }
\; = \; e^{ i \oint_{C} {\bf A \cdot dl } } \; = \;
\; = \; e^{ i \int_S {\bf B \cdot n } \, dA } \;\; ,
\eeq
with $C$ the slightly larger path encircling the two loops.
For a closed surface tiled with many contiguous infinitesimal closed loops
the last expression evaluates to $1$, due to the
divergence theorem. 
In the lattice gravity case the discrete analog of this last result
is considerably more involved, and ultimately represents the (exact) lattice
analog of the contracted Bianchi identities \cite{bianchi}.
An example of a closed surface tiled with parallel transport polygons
(here chosen for simplicity to be triangles) is shown in Fig. 2.

We can now re-examine the question, left open earlier in this Section,
of the value for the quantity $n_d$ appearing in Eq.~(\ref{eq:nd}).
This last quantity counts the number of polygons
needed to 
obtain a closed surface around a hinge, in the framework
of the strong coupling expansion for the curvature correlation function. 
For concreteness, we will consider a simplicial lattice built up
of $d$-dimensional hypercubes divided up into simplices, as
originally discussed in \cite{rowi,hw84} in the four-dimensional case,
although similar considerations should equally apply to other semi-regular
$d$-dimensional lattices as well.
Simply put the issue is then: how many polygons does it take to form the
smallest closed surface attached to two hinges, separated
from each other by $l$ lattice steps?

First let us consider a slightly simpler case, namely the smallest
non-trivial closed surface made out of elementary parallel
transport loops, and built around a {\it single} given hinge. 
In the four-dimensional hypercubic lattice the number of triangles per edge
is either 14 (for the coordinate edges and the hyperdiagonal) or 8 (for
the body and face diagonals).
For a $d$-dimensional lattice, one needs the number of 
$(d-2)$-simplices on each $(d-3)$-simplex.
This again is 14 for some $(d-3)$-simplices, and somewhat less for others.
For example, using the binary notation for the vertices as in \cite{rowi},
if the vertices of the 
$(d-3)$-simplex are taken to be $(0,0,0...), (1,0,0,...), (1,1,0,...), ...$
up to the vertex with $(d-3)$ $1$s followed by $3$ $0$s, then the number 
of $(d-2)$-simplices hinging on this, in the forward direction will be the
same as the number of ways of inserting $1$s in the $3$ remaining places 
with $0$s, which is $7$. 
There will be the same number of $(d-2)$-simplices in the backward direction.
Thus for a typical $(d-3)$-simplex, one needs 14 polygons to form
a closed surface.

The next step then involves considering the minimal closed surface 
connecting {\it two} hinges separated by $l$ lattice steps.
If one is trying to connect two polygonal half-spheres with what
resembles a closed tube, 
one needs to take a path through $(d-2)$-simplices connecting the 
$(d-3)$-simplices at the centers of the half-spheres. 
Suppose the path goes 
through $l$ $(d-2)$-simplices, then the tube will consist of 26 (from the 
ends) plus $12(l-1)$ polygons = $12 l + 14$. 
One noteworthy aspect of this result is that it gives a large power of $k$,
namely $n_d \sim 12$ in the notation of Eq.~(\ref{eq:nd}),
but note that at the same time the power does not grow with $d$.

In the extreme strong coupling limit this then gives, from Eqs.~(\ref{eq:kxi})
and (\ref{eq:nd}),
\beq
\xi \; \mathrel{\mathop\sim_{ k \rightarrow 0}} \; 
{ l_0 \over \vert \log k^{12} \vert }  \; + \; \dots \;\; ,
\label{eq:xi0}
\eeq
where the corrections (indicated here by the dots) arise from surfaces
which are not minimal, i.e. deformations of the original
minimal surface obtained by adding polygonal outgrowths to it, and
therefore involving additional powers of $k$.

\subsection{Random surfaces and the value of the universal exponent $\nu$}

In general for $k$ not too small the random surface spanned by the
parallel transport polygons will have a rather complex shape.
The systematic counting of such surfaces is a rather challenging task,
say compared to a
regular hypercubic lattice, in view of the simplicial nature of the 
underlying lattice geometry.
When discussing the average scalar curvature, given by the
expectation value of $\delta_d \, V_{d-2}$,
such a surface will
be anchored on a given polyhedral loop, whereas when
considering the correlation function of Eqs.~(\ref{eq:corr}) and
(\ref{eq:corrconn}) it will be anchored on two such parallel
transport polygons, separated from each other by some fixed distance
\footnote{
One might worry that the effects of large strong coupling
fluctuations in the ${\bf R}$ matrices might lead to a phenomenon similar
to confinement in non-Abelian lattice gauge 
theories \cite{wilson-lgt,frampton} .
That this is most likely not the case can be seen from the fact
that the analog of the Wilson loop $W(\Gamma)$ 
(defined here as a path ordered exponential
of the affine connection $\Gamma_{\mu\nu}^{\lambda}$ around a closed
loop) does {\it not} give the static gravitational potential.
The potential is instead determined from the correlation of
(exponentials of) geodesic line segments, as in
$ \exp \left [ - \mu_0 \int d \tau \sqrt{ \textstyle g_{\mu\nu} (x)
{d x^{\mu} \over d \tau} {d x^{\nu} \over d \tau} \displaystyle } \;
\right ] $, where $\mu_0$ is the mass of the heavy source,
as discussed already in some detail in \cite{moda,lines}.
The expected decay of near-planar Wilson loops with area $A$, 
$W(\Gamma) \sim \exp ( 
\int_S R^{\, \cdot}_{\;\; \cdot \, \mu\nu} A^{\mu\nu}_{\Gamma} )
\sim \exp ( - A / \xi^2 )$ \cite{peskin}, 
where $A$ is the minimal area spanned by
the loop, gives instead the magnitude of
the large scale, averaged curvature, operationally determined by
the process of parallel-transporting test vectors around very large loops,
and which therefore is of order $R \sim 1 / \xi^2 $.  
}.

As one approaches the critical point, $k \rightarrow k_c$, one is interested
in random surfaces which are of very large extent.
Let $n_p$ be the number of polygons in the surface, and set $n_p = T^2$
since after all one is describing a surface.
The critical point then naturally corresponds to the appearance
of surfaces of infinite extent,
\beq
n_p \; = \; T^{2} \; 
\sim \; 
{ 1 \over  k_c - k }  \; \rightarrow \; \infty \;\; .
\label{eq:T}
\eeq
A legitimate parallel is to the simpler case of scalar field theories, 
where random walks of length $T$ describing particle paths become of
infinite extent at the critical point, situated where the inverse of
the (renormalized) mass $\xi=m^{-1}$, expressed in units of the 
ultraviolet cutoff, diverges \cite{feypath,symrw,haus,gross-dtrs,phi4}.

In the present case of polygonal random surfaces, one can provide the
following concise argument in support of the identification
in Eq.~(\ref{eq:T}).
First approximate the discrete sums over $n$, as they appear for example
in the strong coupling expansion for the average curvature, 
Eq.~(\ref{eq:rseries}) or
its correlation, Eq.~(\ref{eq:corr}), by continuous integrals over
areas
\beq
\sum_{n=0}^\infty \, c_n \, \left ( { k \over k_c } \right )^n  \;
\rightarrow \;
\int_0^\infty dA \, A^{\gamma-1} \left ( { k \over k_c } \right )^A
\; = \; 
\Gamma ( \gamma ) \, \left ( \log { k_c \over k } \right )^{-\gamma}
\;\; ,
\label{eq:areasum}
\eeq
where $A \equiv T^2$ is the area of a given surface.
The $A^{\gamma-1}$ term can be regarded as counting the
multiplicity of the surface (its entropy, in statistical mechanics terms).
The exponent $\gamma$ depends on the specific quantity one is looking at.
For the average curvature one has $\gamma=-\delta$, while for its derivative,
the curvature fluctuation (the curvature correlation function at
zero momentum), one expects $\gamma=1-\delta$.
The same type of singularity is of course obtained from the original
series in Eq.~(\ref{eq:areasum}), if one assumes for the coefficients $c_n$
\beq
c_n \; \sim \; { - \gamma \choose n }
\; \mathrel{\mathop\sim_{ n \rightarrow \infty }} \;
{ \Gamma ( 1 - \gamma ) \sin \pi (n+ \gamma) \over \pi } \; n^{\gamma-1} \,
\left ( 1 \, - \, { \gamma \, ( 1 - \gamma ) \over 2 \, n } 
\, + \, \dots \right )
\;\; ,
\eeq
which in retrospect explains the appearance of the factor 
$A^{\gamma-1}$ in Eq.~(\ref{eq:areasum}).
In the last step we have used the well-known asymptotic expansion for the
binomial coefficient ${ - \gamma \choose n } $ for large $n$. 
Although we know its value exactly, the integral in 
Eq.~(\ref{eq:areasum}) can also be evaluated by
standard saddle point methods.
The saddle point is located at
\beq
A \; = \; { ( \gamma - 1 ) \over \log { k_c \over k } } 
\; \mathrel{\mathop\sim_{ k \rightarrow k_c }} \;
{ ( \gamma - 1 ) \, k_c \over k_c - k } \;\; .
\eeq
Carried further, the saddle point method then leads to an
approximation to the exact result for the quantity in
Eq.~(\ref{eq:areasum}), namely
\beq
e^{1-\gamma} \, ( \gamma - 1 )^{\gamma-1} \, \sqrt{2 \pi (\gamma-1) }
\left ( \log { k_c \over k } \right )^{-\gamma} \;\; ,
\label{eq:saddle}
\eeq
which agrees with the answer given above,
up to an irrelevant overall multiplicative factor.
From this discussion one then concludes that close to the critical point
very large areas dominate, as claimed in Eq.~(\ref{eq:T}).

Furthermore, one would expect that the universal geometric scaling properties
of such a (closed) surface would not depend on its short distance
details, such as whether it is constructed out
of say triangles or more complex polygons.
In general excluded volume effects at finite $d$ will provide constraints on
the detailed geometry of the surface, but as $d \rightarrow \infty$ these
constraints can presumably be neglected and one is dealing then with a
more or less unconstrained random surface.
In the following we will assume that this is indeed the case, and that
no special pathologies arise, such as the collapse of the
random surface into narrow tube-like, lower dimensional geometric
configurations.
Then in the large $d$ limit the problem simplifies considerably.

Following \cite{gross-dtrs}, one can define the partition function
for such an ensemble of unconstrained random surfaces as
\beq
Z_{RS} \; = \; \int \prod_{n,m=1}^T \, d^d X_{n,m} \, \exp \left [
- \, \beta \, \sum_{\Delta} \, A_{\Delta} ( {\bf X} ) \right ] \;\; ,
\eeq
where the integral is over $d$-component vectors ${\bf X}_{n,m}$
defined on two-dimensional triangular lattice sites, with
sites labeled here by integers $n$ and $m$.
Up to a multiplicative constant, the term appearing in the exponent
is just the total area of the surface, written as a sum
of individual triangle areas.
Introducing the induced two-dimensional metric tensor on each triangle
allows one to recast the above partition function in the form
of a two-dimensional massless field theory,
which in a more compact continuum notation now reads
\beq
Z_{RS} \; = \; {\rm const. } \, \int [d \lambda] [d g] [d {\bf X} ] \,
\exp \left [
i \int d^2 x  \sqrt{g} \, \lambda^{ab} \, ( g_{ab} \, - \, G_{ab} )
\, - \, \beta \, \int d^2 x \sqrt{g} \, \right ] \;\; ,
\eeq
with $G_{ab} = \partial_a {\bf X} \cdot  \partial_b {\bf X} $.
The above action is now quadratic in the free massless ${\bf X}$-fields,
whose propagator involves $\lambda$-dependent weights.
We note that in the original gravitational context,
the introduction of the coordinate vectors ${\bf X}(x)$
for describing the random surface spanned by polygons, 
originally embedded in a fluctuating {\it curved} geometry,
would seem plausible in view of the fact that as one approaches the
critical point the expectation value of the scalar curvature
does indeed go to zero \cite{critical}.

As shown in \cite{gross-dtrs},
the overall {\it size} of the random surface, as embedded in the
original $d$-dimensional space and suitably defined in the discrete case as
\beq
< {\bf X}^2 > \; = \; { 1 \over T^2 } \, \sum_{n,m=1}^T \, {\bf
  X}_{n,m}^2 \;\; , 
\eeq 
is then immediately obtained from the free field infrared behavior of ${\bf X}$
as $< {\bf X}^2 > \sim \int_{1/T} \, d^2 p / p^2 \sim \log T $.
Thus the mean square size of the surface increases logarithmically with
the intrinsic area of the surface.
This last result is usually interpreted as the statement that an
unconstrained random surface has infinite fractal (or Hausdorff) dimension.
Although made of very many triangles (or polygons), the random
surface remains quite compact in overall size, as viewed
from the original embedding space.
In a sense, an unconstrained random surface is a much more compact object
than an unconstrained random walk, for which $< {\bf X}^2 > \sim T $.
Identifying the size of the random surface with the gravitational
correlation length $\xi$ then gives
\beq
\xi \; \sim \; \sqrt{ \log T} 
\; \mathrel{\mathop\sim_{ k \rightarrow k_c }} \;
\vert \log ( k_c - k ) \vert^{1/2} \;\; .
\label{eq:xilog}
\eeq
From the definition of the exponent $\nu$, namely
$\xi \sim (k_c - k)^{-\nu}$, the above result then implies
$\nu = 0$ (i.e. a weak logarithmic singularity)
\footnote{
In four dimensions one finds for lattice quantum gravity
$\nu \approx 1/3 $ instead \cite{critical,ttmodes}.
}.
We note that the previous result for $\xi$ in Eq.~(\ref{eq:xi0}) only applied
to the extreme strong coupling limit $ k \rightarrow 0$.

Let us discuss next what the implications of this last result might be.
As already outlined in Refs. \cite{lines,ttmodes,critical}, 
the exponent $\nu$ determines the universal renormalization group
evolution of the dimensionless coupling 
$\tilde G \equiv G \, \lambda^{(d-2)/d}$ in the vicinity of the
ultraviolet fixed point.
In particular, if one defines the dimensionless function $F(\tilde G)$ via
$m \equiv \xi^{-1} = \Lambda \, F(\tilde G)$, where $\Lambda$ is
the ultraviolet cutoff (the inverse lattice spacing), then
by differentiation of the renormalization group invariant quantity $m$,
$ \Lambda \, {d \over d \, \Lambda} \, m ( \Lambda, \tilde G (\Lambda) ) =0 $, 
one immediately obtains the Callan-Symanzik beta 
function $\beta ( \tilde G )$ \cite{frampton}.
From the definition
\beq
\Lambda \, { d  \over d \, \Lambda } \, \tilde G ( \Lambda ) \; = \; 
\beta ( \tilde G ( \Lambda ) ) \;\; ,
\label{eq:betaf}
\eeq
one gets an equivalent form for the beta function in terms of the
function $F(\tilde G)$ introduced above, namely
\beq
\beta ( \tilde G ) \; = \; - \, { F(\tilde G) \over 
\partial F(\tilde G) / \partial \tilde G } \;\; .
\eeq
The generic beta function equation, determining
the scale evolution of the coupling (obtained from Eq.~(\ref{eq:betaf}),
and identical in form to it),
\beq
\mu \, { d  \over d \, \mu } \, \tilde G ( \mu ) \; = \; 
\beta ( \tilde G ( \mu ) ) \;\; ,
\label{eq:betamu}
\eeq
can then be integrated in the vicinity of the fixed point,
leading to a definite relationship between the relevant coupling 
$\tilde G$, the renormalization group invariant (cutoff independent)
quantity $m=1/\xi$, and an arbitrary sliding scale $\mu=1/r$.  
Up to scales of order $\xi$, it determines the universal running of
$\tilde G$, which will give rise to macroscopic effects provided the 
non-perturbative scale $\xi$ is very large.
In \cite{cosm,ttmodes} this scale was naturally identified with the
scaled cosmological
constant, which here would correspond to the ratio $\lambda / G$.
The result of Eq.~(\ref{eq:xilog}) then corresponds to the limiting
case $\nu \rightarrow 0$.
In the language of Refs. \cite{cosm,ttmodes}, it leads in the vicinity
of the fixed point to an exponentially small (for $r/\xi \rightarrow 0$) 
renormalization-group running of $\tilde G ( \mu )$ or $\tilde G ( r )$,
namely
\beq
\tilde G ( r ) \, - \, \tilde G_c 
\; \mathrel{\mathop\sim_{ \tilde G \rightarrow \tilde G_c }} \; 
e^{- c \, ( \xi / r )^2 } \;\; .
\label{eq:grun}
\eeq

All of the above was in the limit of infinite dimension.
In Ref. \cite{ttmodes} it was suggested, based on a simple geometric
argument, that $\nu = 1 / (d-1)$ for large $d$.
Moreover, for the lattice theory in finite dimensions one finds no phase
transition in $d=2$ \cite{hw2d}, $\nu \approx 0.60$ in $d=3$ \cite{hw3d}
and $\nu \approx 0.33$ in $d=4$ \cite{critical,ttmodes},
which then leads to the (almost constant) sequence 
$(d-2) \nu$ = $1$, $0.60$ and $0.66$ in the three cases respectively.
After interpolating this last series of values with a quadratic
polynomial in $1/d$, one obtains $\nu \approx 1.9/d$ for large $d$.
On the other hand in Ref. \cite{litim} the value  $\nu = 1/2d$ was obtained
in the same limit with a Wilson-type continuum renormalization group approach,
in which a momentum space slicing technique is combined
with a truncation to the Einstein-Hilbert action and a cosmological term.
It seems that in either case our analytical results for the large $d$
limit are consistent with, and to some extent corroborate,
these previous findings.
For completeness let us mention here that in the extreme opposite case,
namely close to two dimensions, one has the by now well-established 
result $\nu = 1 / (d-2) + O( (d-2)^0 ) $ \cite{epsilon,epsilon1}.

It is of interest to contrast the result $\nu \sim 0$ for gravity in
large dimensions with what one finds for scalar \cite{wilson,fisher}
and gauge \cite{DPS} fields, in the same limit $d=\infty$.
Known results, and what we have found here so far, can be
combined and summarized as follows
\bea
& {\rm scalar \; field } \;\;\;\;\;\;\;\;\;\;\;\; & \nu \; = \; \half 
\nonumber \\
& {\rm lattice \; gauge \; field } \;\; & \nu \; = \; \quarter
\nonumber \\
& {\rm lattice \; gravity } \;\;\;\;\;\;\; & \nu \; = \; 0 \;\; . 
\eea
The first rather well-known result is re-derived in Appendix A.
The second one, obtained for non-Abelian gauge theories at large $d$, 
is recalled in Appendix B.
It should be regarded as encouraging that the new value obtained here,
namely $\nu=0$ for gravitation, appears to some extent to be consistent with
the general trend observed for lower spin, at least at infinite dimension.

As far as $1/d$ corrections are concerned, 
the result obtained previously in this section hinge on the crucial assumption
that the random surface is non-interacting, in other words
that any self-intersection or folding of the surface does not carry
additional statistical weights.
This is similar to an unconstrained random walk, where the effects
of path intersection and backtracking are neglected.
While these assumptions seem legitimate at infinite $d$ (since
there are infinitely many orthogonal dimensions to move into), they
are no longer valid at finite $d$.
As a result, the problem becomes much more complex, and one
expects that $\nu$ will then no longer be equal to zero.
Indeed in four dimensions $\nu \approx 1/3$ \cite{critical}.
In the much simpler random walk case, a systematic expansion can be developed,
leading for $n$ intersections to an effective $\phi^{2(n-1)}$ interaction
for the scalar field associated with the random walk.
Unfortunately in the gravitational case it is much less clear how to
develop such a systematic expansion.


\vskip 30pt
\newsection{The Continuum Case}
\hspace*{\parindent}

For quantum gravity formulated in dimensions greater than four there 
are a number of natural questions that come to mind.
Are there any special dimensions for gravity?
How do the Feynman rules depend on $d$?
What does continuum gravity look like in large dimensions?
Before discussing the gravitational case, it might be useful
to examine and contrast the somewhat simpler cases of 
scalar and vector (gauge) theories.

\subsection{Special values of $d$ in field theories}

In {\it scalar} field theories the special role of dimension four
is easily brought out by writing the action, simply using dimensional
arguments, as
\beq
S \; = \; { 1\over 2} \int d^d x \, \left [ \,
( \, \partial_{\mu} \phi (x) \, )^2 \, - \, m_0^2 \, \phi^2 (x) \; \right ]
\; - \;  { \lambda_0 \over \Lambda^{d-4} } \int d^d x \,
\phi (x)^4 \;\; ,
\label{eq:phi4}
\eeq
where $\Lambda$ is the ultraviolet cutoff, $\lambda_0$ the bare
self-coupling, $m_0$ the bare mass, and with the fields having
canonical dimension $m^{(d-2)/2}$.
The self-coupling is dimensionless only in dimension four, and
above that the model is described in the long-distance,
infrared limit by a free field \cite{phi4}.
The interaction term is {\it relevant} for $d<4$, and
{\it irrelevant} above $d=4$.
In particular for any $d>4$ one can prove that the correlation
length exponent $\nu$ equals one half, the free field value \cite{wilson,fisher}. 
The long distance, infrared behavior is the same as for a free field.

In the case of $SU(N)$ non-Abelian {\it gauge theories}
one has that the coupling is, again, dimensionless only
in four dimensions, a well-known signature of perturbative
renormalizability.
Above four dimensions purely dimensional arguments
indicate the appearance of a non-trivial ultraviolet fixed point
(a zero of the Callan-Symanzik $\beta(g)$ function) close to the origin,
\beq
\beta (g ) \; = \; (d-4) \, g \, - \, \beta_0 \, g^3 \, + \, \dots
\;\; ,
\eeq
with a non-trivial fixed point at $g^2_c = (d-4) / \beta_0 + O( (d-4)^2 )$
separating what is believed to be a Coulomb, non-confining phase, from
the confining phase known to exist for sufficiently strong 
coupling \cite{wilson-lgt}. 
Since the theory is not perturbatively renormalizable
above four dimensions, the analysis of either phase is rather problematic
in the continuum.
The transition is characterized by non-trivial critical exponents,
and the Green's functions in the scaling region correspond to an
interacting theory, which can only be reconstructed in the 
Coulomb phase $g< g_c$ as an expansion in $\e=d-4$ \cite{parisi}.

One might wonder if anything special happens in dimensions
$d>4$, beyond what has just been discussed.
In $SU(N)$ Yang-Mills with (Euclidean) classical action 
\beq
I_{cl} \; = \; { 1 \over g^2 \, N } \, \int d^d x \, 
{ \textstyle {1 \over 4} \displaystyle } \, \tr F_{\mu\nu}^2 \;\; ,
\eeq
one has to one loop for the divergent part of the effective action
\beq
\Gamma_{div}^{(1)} \; = \; {1 \over 4-d} \, { 26-d \over 3 } \,
{ g^2 \, N \over 16 \pi^2 } \, I_{cl} \;\; ,
\eeq
which vanishes in $d=26$, and to two loops
\beq
\Gamma_{div}^{(2)} \; = \; {1 \over 4-d} \, { 34 \over 3 } \,
\left ( { g^2 \, N \over 16 \pi^2 } \right )^2 \, I_{cl}
\eeq
\cite{caswell}.
One would be hard pressed though to conclude that the above
results suggest anything dramatic might happen at $d=26$
in the Yang-Mills case, as the change of sign in the one loop
divergence is still counteracted by the two-(and higher-) loop
terms for sufficiently large $g^2$.
It seems in general that the structure of the continuum
theory at large $d$ remains
quite complicated and possibly still not amenable to a perturbative treatment.

On the lattice on the other hand the presence of a phase transition
has been clearly established in the large $d$ limit, in fact largely
irrespective of the specific choice of continuous symmetry group \cite{DPS}.
For the group $SU(N)$ a critical point in $g$ appears at 
$ 2 d \, (2 N / g^2)^4 = const$,
(with the constant depending of the specific choice of $N$),
and with an exponent at the transition given by $\nu=1/4$ \cite{DPS}.
But it seems that finding such a transition critically
hinges on using non-perturbative methods, which allow one
to explore the strong coupling regime,
and in particular the existence of two physically distinct phases.

In the case of {\it gravity}, the expression analogous to
Eq.~(\ref{eq:phi4}) is
\beq
\lambda \, \int d^d x \, {\textstyle {\sqrt{g}} \displaystyle}
\, - \, { 1 \over 16 \, \pi \, G} \, 
\int d^d x \, {\textstyle {\sqrt{g}} \displaystyle} \, R 
\, + \, 
{ \alpha_0 \over \Lambda^{4-d} } \, 
\int d^d x \, {\textstyle {\sqrt{g}} \displaystyle} \, R_{\mu\nu}
R^{\mu\nu} 
\, + \, 
{ \beta_0 \over \Lambda^{4-d} } \, 
\int d^d x \, {\textstyle {\sqrt{g}} \displaystyle} \, R^2 
\, + \, \cdots \;\; ,
\label{eq:highder}
\eeq
which shows the suppression of the curvature squared terms in the
infrared region, by factors $O(1/\Lambda^2)$ when
compared to the Einstein term, whose coefficient also involves
a dimensionful quantity, namely $\Lambda^{d-2} / (16 \, \pi \, G_0) $
(here $\alpha_0$ and $\beta_0$, as well as $G_0 \equiv \Lambda^{d-2} G$, 
are taken to be dimensionless couplings)
\footnote{
Adding curvature squared terms to the bare action cures the 
perturbative non-renormalizability problem, but raises new issues
related to unitarity \cite{hdqg}.
Curvature squared terms are expected to play important roles
at very short distances, comparable to the cutoff scale,
where fluctuations in the curvature
can become of order $\sim \Lambda^2 / G_0$.
}.
It then seems legitimate to ask if there are any special dimensions
for gravity, in particular above $d=4$.
As already mentioned in the Introduction,
one has $d(d+1)/2$ independent components of the
metric in $d$ dimensions, and the same number of algebraically independent
components of the Ricci tensor appearing in the field equations. 
The contracted Bianchi identities reduce the
count by $d$, and so does general coordinate invariance,
leaving $d(d-3)/2$ physical gravitational
degrees of freedom in $d$ dimensions.
As a result, the number of physical degrees of freedom of the gravitational
field grows rather rapidly (quadratically) with the number of dimensions.

The first step is naturally to examine tree level gravity, where all
loop (quantum) effects are neglected \cite{feylec,deser04,deser3d}.
Then in the non-relativistic, static limit gravitational interactions are
described by
\beq
I_2 [T]  \; = \; - \, {\kappa^2 \over 2 } \, \int d^d x \,
\left [ T_{\mu\nu} \, \Box^{-1} \, T^{\mu\nu} \, - \, 
(d-2)^{-1} \, T_{\mu}^{\;\;\mu} \, \Box^{-1} \, T_{\nu}^{\;\;\nu} \right ]
\nonumber \\
\; \rightarrow \; 
- \, {d-3 \over d-2 } \, {\kappa^2 \over 2 } \, \int d^{d-1} x \,
T^{00} \, {\cal G} \, T^{00} ,
\eeq
where the Green's function $\cal{G}$ is the static limit of $1 / \Box$,
and $\kappa^2 = 16 \pi G$.
The above result then incorporates at least two well-known facts,
namely that there are no Newtonian forces in $d$=2+1 dimensions,
and that the Einstein tensor vanishes
identically in $d$=1+1 dimensions.
But nothing particularly noteworthy seems to happen, at least at tree level,
above $d=3$.
At the same time, four spacetime dimensions is known to be the lowest dimension
for which Ricci flatness does not imply the vanishing of the gravitational
field, $R_{\mu\nu\lambda\sigma}=0$, and therefore the first dimension
to allow for gravitational waves and their quantum counterparts, gravitons.
The tree level {\it static} gravitational potential above $d>3$ is simply
obtained by Fourier transform using
\beq
\int d^d x \, { 1 \over k^2 } \, e^{i \, k \cdot x } \; = \;
{ \Gamma \left ( {d - 2 \over 2} \right ) \over 4 \, \pi^{d/2} \,
( x^2 )^ { d /2 - 1} }
\eeq
and therefore implies 
$\int d^{d-1} {\bf x} \, e^{i {\bf k} \cdot {\bf x} } / {\bf k}^2 $
$\sim 1 / r^{d-3}$.

When quantum loop effects are turned on \cite{thooft,deser},
one finds that the one-loop divergence,
proportional to curvature squared terms, vanishes on shell,
\beq
\Gamma_{div}^{(1)} \; = \; {1 \over 4-d} \, { \hbar \over 16 \pi^2 } 
\int d^4 x \sqrt{g} \; \left ( 
{ 7 \over 20 } \, R_{\mu\nu} \, R^{\mu\nu} \, + \, { 1 \over 120 } \, R^2 
\right ) \;\; ,
\eeq
using the well known result  
$R_{\mu\nu\rho\sigma} \, R^{\mu\nu\rho\sigma} \; = \;
- \, R^2 \, + \, 4 \, R_{\mu\nu} \, R^{\mu\nu} 
\, + \, {\rm total \; derivative}$ to eliminate Riemann squared
terms.
The complete set of one loop divergences, computed using the heat
kernel expansion and zeta function regularization close to four
dimensions, can be found in the comprehensive
review cited in \cite{hawking}, and further references therein.
At two loops it was shown some time ago \cite{sagnotti,vandeven}
that there is a non-removable on-shell two-loop $R^3$-type divergence
\beq
\Gamma_{div}^{(2)} \; = \; {1 \over 4-d} \, { 209 \over 2880} \,
{ \hbar^2 \, G \over (16 \pi^2)^2 } 
\int d^4 x \sqrt{g} \; 
R_{\mu\nu}^{\;\;\;\;\rho\sigma} \, R_{\rho\sigma}^{\;\;\;\;\kappa\lambda} 
\, R_{\kappa\lambda}^{\;\;\;\;\mu\nu} \;\; .
\eeq
In the last quoted reference it is argued that in the above expression
the $209$ arises from $11 \times 19$, with the factor of 11 coming
from $(26-d)/2 $, as expected from closed string theory \cite{vandeven}.
Thus the latter divergence might vanish again at $d=26$, but it is not
expected that the same will happen at higher loops.

Recent two-loop results based on the $2+\epsilon$ 
expansion for gravity with a cosmological constant \cite{epsilon},
inspired by the $2+\epsilon$ of other, simpler
field theory models \cite{wilseps,sigma,grossneveu},
show the appearance of a non-trivial ultraviolet fixed point in the
$G$ beta function above two dimensions,
\beq
\beta (G) \, = \, (d-2) \, G \, - \, { 2 \over 3 } (25- n_f) \, G^2 \,
- \, { 20 \over 3 } (25- n_f) \, G^3 \, + \cdots \;\; ,
\label{eq:betaeps}
\eeq
(for $n_f$ massless real scalar fields minimally coupled to gravity).
They could be possibly relevant as
a first crude approximation to the four-dimensional theory (to the extent that
they represent a manifestly gauge invariant resummation of those diagrams
which can be regarded as dominant close to two dimensions).
But unfortunately they can hardly be
thought as useful in the limit $d \rightarrow \infty$, especially
in view of the fact that the Borel summability in $\epsilon=d-2$ 
\cite{hikami,kleinert} of such an
expansion still remains a largely open question.

\subsection{Feynman rules in $d$ dimensions}

A direct examination of the Feynman rules for continuum gravity at large
$d$ indeed reveals the occurrence of some degree of simplification.
But first we should clarify our
conventions and notation for this section, which are taken from \cite{pot},
and where one expands around the flat Minkowski space-time metric, with
signature given by
$\eta_{\mu\nu}={\rm diag}(1,-1,-1,-1, \dots)$.
The Einstein-Hilbert action in $d$ dimensions is then given by
\beq
S_{\rm E} \; = \; + {1\over 16\pi G} \int d^d x \,
{\textstyle \sqrt{-g(x)} \displaystyle} \, R(x)\, \;\; ,
\eeq
with $g(x)={\rm det}(g_{\mu\nu})$ and $R$ the scalar curvature
(it will also be assumed in the following that the bare cosmological
constant is zero). 
Furthermore the coupling of gravity to scalar particles of mass $m$
is described by the action
\beq
S_{\rm m} \; = \; { 1\over 2} \int d^d x \;
{\textstyle \sqrt{-g(x)} \displaystyle} \,
\left [ \; g^{\mu \nu} (x) \,
\partial_{\mu} \phi (x) \, \partial_{\nu} \phi (x) \; - \;
m^2 \, \phi^2 (x) \; \right ] \;\; .
\eeq
Usually in perturbation theory the metric $g_{\mu\nu} (x) $
is expanded around the flat metric $\eta_{\mu\nu}$ \cite{thooft},
by writing
\beq
g_{\mu\nu}(x) \; = \; \eta_{\mu\nu} + \kappa \; {\tilde h}_{\mu\nu}(x)
\;\; ,
\eeq
with $\kappa^2=32 \pi G$.
In the harmonic (de Donder) gauge the graviton propagator is then given by
\beq
D_{\mu\nu \rho\sigma}(p) \; = \; 
{ i \over 2} \; { \eta_{\mu\rho}\eta_{\nu\sigma}+
\eta_{\mu\sigma}\eta_{\nu\rho}- {2 \over d-2} \eta_{\mu\nu}\eta_{\rho\sigma}
\over p^2 + i \e } \;\; ,
\label{eq:prop}
\eeq
which suggests that the conformal mode contribution
might go away as $ d \rightarrow \infty$.
But further thought reveals that this conclusion might perhaps be
fallacious, as a different type of expansions seem to lead to slightly
different conclusions.

If one follows the method of reference \cite{fpgrav,capper}, then one defines
the small fluctuation graviton field $h_{\mu\nu}(x)$ instead via
\beq
g^{\mu\nu}(x)
{\textstyle \sqrt{-g(x)} \displaystyle} 
\; = \; \eta^{\mu\nu} + \kappa \; h^{\mu\nu}(x) \;\; .
\eeq
One advantage of this expansion over the previous one is that it leads
to considerably simpler Feynman rules, both for the graviton vertices
and for the scalar-graviton vertices.
A gauge fixing term can then be added \cite{fp,ft}, for example of the form 
\beq
{ 1 \over \kappa^2 }
\left ( \partial_\mu
{\textstyle \sqrt{-g(x)} \displaystyle} \, g^{\mu\nu} \right )^2 \;\; ,
\eeq
as again used in \cite{capper}.
The bare graviton propagator is then given simply by
\beq
D_{\mu\nu \rho\sigma}(p) \; = \; 
{ i \over 2} \; { \eta_{\mu\rho}\eta_{\nu\sigma}+
\eta_{\mu\sigma}\eta_{\nu\rho}-\eta_{\mu\nu}\eta_{\rho\sigma}
\over p^2 + i \e } ,
\eeq
whose structure is now unaffected by the limit $ d \rightarrow \infty$.
Thus with the latter definition for
the gravitational field, there are no factors of $1/(d-2)$ for the graviton
propagator in $d$ dimensions; such factors appear instead in the expressions
for the Feynman rules for the vertices.
For the three-graviton and two ghost-graviton vertex the
relevant expressions are quite complicated.
The three-graviton vertex can be written as 
\bea
&& U(q_1,q_2,q_3)_{\a_1 \b_1, \a_2 \b_2, \a_3 \b_3} = \nonumber \\
&& - i { \kappa \over 2 } \Bigl [
q^2_{(\a_1} q^3_{\b_1)} 
\left (
2 \et_{\a_2(\a_3} \et_{\b_3)\b_2} - {\textstyle {2 \over d-2} \displaystyle}
\et_{\a_2\b_2} \et_{\a_3\b_3}
\right ) \nonumber \\
&& \;\;\;\;\;\; + 
q^1_{(\a_2} q^3_{\b_2)} 
\left (
2 \et_{\a_1(\a_3} \et_{\b_3)\b_1} - {\textstyle {2 \over d-2} \displaystyle}
\et_{\a_1\b_1} \et_{\a_3\b_3}
\right ) \nonumber \\
&& \;\;\;\;\;\; +
q^1_{(\a_3} q^2_{\b_3)} 
\left (
2 \et_{\a_1(\a_2} \et_{\b_2)\b_1} - {\textstyle {2 \over d-2} \displaystyle}
\et_{\a_1\b_1} \et_{\a_2\b_2}
\right )  \nonumber \\
&& \;\;\;\;\;\; +
2 q^3_{(\a_2} \et_{\b_2)(\a_1} \et_{\b_1)(\a_3}  q^2_{\b_3)} +
2 q^1_{(\a_3} \et_{\b_3)(\a_2} \et_{\b_2)(\a_1}  q^3_{\b_1)} +
2 q^2_{(\a_1} \et_{\b_1)(\a_3} \et_{\b_3)(\a_2}  q^1_{\b_2)}  \nonumber \\
&& \;\;\;\;\;\; +
q^2 \cdot q^3  \left ( 
{\textstyle {2 \over d-2} \displaystyle} \et_{\a_1(\a_2} \et_{\b_2)\b_1} \et_{\a_3\b_3} +
{\textstyle {2 \over d-2} \displaystyle} \et_{\a_1(\a_3} \et_{\b_3)\b_1} \et_{\a_2\b_2} -
2 \et_{\a_1(\a_2} \et_{\b_2)(\a_3} \et_{\b_3)\b_1} \right ) \nonumber \\
&& \;\;\;\;\;\; +
q^1 \cdot q^3  \left ( 
{\textstyle {2 \over d-2} \displaystyle} \et_{\a_2(\a_1} \et_{\b_1)\b_2} \et_{\a_3\b_3} +
{\textstyle {2 \over d-2} \displaystyle} \et_{\a_2(\a_3} \et_{\b_3)\b_2} \et_{\a_1\b_1} -
2 \et_{\a_2(\a_1} \et_{\b_1)(\a_3} \et_{\b_3)\b_2} \right ) \nonumber \\
&& \;\;\;\;\;\; +
q^1 \cdot q^2  \left ( 
{\textstyle {2 \over d-2} \displaystyle} \et_{\a_3(\a_1} \et_{\b_1)\b_3} \et_{\a_2\b_2} +
{\textstyle {2 \over d-2} \displaystyle} \et_{\a_3(\a_2} \et_{\b_2)\b_3} \et_{\a_1\b_1} -
2 \et_{\a_3(\a_1} \et_{\b_1)(\a_2} \et_{\b_2)\b_3} \right )
\Bigr ] \;\; . \nonumber \\
\eea
Again one notes that some terms become negligible as $d \rightarrow \infty$,
but the remaining ones can have either sign, giving rise
to non-trivial cancellations even for large $d$.
The ghost-graviton vertex is given by
\beq
V(k_1,k_2,k_3)_{\a \b, \l \m} = i \kappa \left [
- \et_{\l(\a} k_{1 \b)} k_{2\m } 
+ \et_{\l \m} k_{2(\a)} k_{3\b)} 
\right ] \;\; ,
\eeq
and the two scalar-one graviton vertex is given by
\beq
{ i \kappa \over 2 } \left ( p_{1 \mu} p_{2 \nu} + p_{1 \nu} p_{2 \mu} -
{ 2 \over d-2} \; m^2 \; \eta_{\mu\nu} \right ) \;\; ,
\eeq
where the $p_1,p_2$ denote the four-momenta of the incoming and outgoing
scalar field, respectively.
Finally the two scalar-two graviton vertex is given by
\beq
{ i \kappa^2 m^2 \over 2(d-2) } \left ( \eta_{\mu \lambda} \eta_{\nu \sigma}
+ \eta_{\mu \sigma} \eta_{\nu \lambda} 
- { 2 \over d-2 } \; \eta_{\mu \nu} \eta_{\lambda \sigma} \right )
\;\; , 
\eeq
where one pair of indices $(\mu,\nu)$ is associated with one graviton line,
and the other pair $(\lambda,\sigma)$ is associated with the other graviton
line.
Again one notices some simplification in the limit $d \rightarrow \infty$.
These rules follow readily from the expansion of the gravitational action to
order $G^{3/2}$ ($\kappa^3$), and of the scalar field action to order
$G$ ($\kappa^2$).

The next step would involve a careful analysis of what the dominant diagrams
are in the large $d$ limit (still keeping in mind the
serious shortcoming of assuming a vanishing bare cosmological
constant), assuming that
such a procedure remains reliable in this limit, in the sense that a
complete resummation can be performed, and that there are no
large non-perturbative, non-analytic contributions.
But it seems so far that in the case of gravity there are conflicting 
claims in the literature \cite{strom1,strom2,bb} as to what 
exactly happens in the continuum as $ d \rightarrow \infty$.

In Refs. \cite{strom1,strom2} a gauge-invariant expansion in $1/d$
was developed for vanishing bare cosmological constant,
considering both the case where the extra dimensions
are non-compact and the case where they are highly compactified.
The observation was made that there are order-by-order (in $1/d$)
cancellations of large numbers of graphs, but the origin of
such cancellation remained a puzzle. 
However, it was found that the leading term of any Green's function
was given by a set of disjoint bubble graphs. 
It was then determined that the graviton propagator
acquires a physical pole near the Planck mass, unfortunately
in a region where the validity of the expansion appears questionable. 
Finally it was claimed that at $d=\infty$ phase-space factors suppress
the Feynman integrations and the theory is therefore finite.

In the recent work of \cite{bb} it is also claimed that a consistent
leading large-$d$ limit exists for the Einstein theory without
cosmological term, and that it can be constructed 
using a sub-class of planar diagrams, which seems somewhat in disagreement
with the class of diagrams identified in the previous references.
It is then found that the large-$d$ quantum
gravity limit is well defined and renormalizable, 
provided the space-time integrations are not extended to the full
$d$-dimensional space-time, in other words if the full space-time
allows for compactified dimensions (the last result does not seem
entirely surprising, as compactifying and shrinking extra dimensions
leads to an effectively lower dimensional theory, with possibly
convergent momentum integrations, depending on how the limits are taken).

But it seems difficult to reconcile the above quoted results with
the fact that a) the perturbative non-renormalizabilty issue
only gets worse in the continuum as one increases the dimension,
and b) that at least close to two dimensions an ultraviolet fixed
point is {\it known} to exist, and somehow completely
fails to show up in the large $d$ diagrammatic treatment.
The more likely scenario is perhaps that the theory remains
perturbatively non-renormalizable even in the large $d$ case, and
therefore just as intractable in the continuum as
the equally difficult large-$d$ Yang-Mills case.
So far the continuum perturbative
diagrammatic treatment has not lead yet to any conclusive predictions
about the behavior of physical gravitational observables 
(such as scale dependence and renormalization of
couplings, nontrivial fixed points, anomalous scaling dimensions etc.),
which makes it difficult to compare with recent non-perturbative 
lattice \cite{critical,ttmodes} and continuum \cite{epsilon,litim}
results in low dimensions, both of which project a rather
different picture.

\vskip 30pt
\newsection{Conclusions}
\hspace*{\parindent}

In this paper we have examined the lattice formulation of quantum gravity
in the large $d$ limit.
Such a line of inquiry was stimulated by the fact that statistical systems
based on local interactions generally tend to simplify considerably in this
limit, where each point is found to be surrounded by a large number of
neighbors, and mean field theory methods apply.
Even when mean field theory does not apply, the hope was that the theory
would simplify significantly, to a point where it could be solved exactly.
In view of the general lack of analytical results, aside from perturbation
theory and some other investigations restricted to low and somewhat
unphysical dimensions,
one would expect such results would help shed new light on the true
non-perturbative ground state of quantum gravity in four dimensions.

While $d=\infty$ at first seems rather remote from the physical case 
$d=4$, one can make the case that the well known $1/N$ expansion of
statistical mechanics system and $SU(N)$ gauge theories (the planar limit)
has lead to some remarkable insights into the finite $N$
structure of these theories, and in some cases even to quantitatively
accurate answers for critical exponents (in the statistical mechanics
context) and specific phenomenological predictions 
(for example in low energy QCD applications).
Indeed more that once it has been argued that
in the case of QCD, based on both theoretical and phenomenolgical 
arguments, that $1/N=0$ is not too remote from the physical case $1/N=1/3$.
In the same spirit, $1/d=0$ might not be as remote as it seems at
first from the real world case of $1/d=1/4$ theory.

In pursuing the $1/d$ expansion for gravity we have followed two 
somewhat complementary approaches. 
In the first approach, various terms which appear
in the lattice gravitational (Regge) action were expanded in powers of $1/d$.
Since the resulting expressions are still rather cumbersome, we resorted
to a combined weak field expansion, perturbing arbitrarily
coordinated lattices built out of nearly equilateral simplices.
The resulting expressions were then evaluated for the cross polytope,
a triangulation of the $d$-dimensional sphere based on the 
dual of a $d$-dimensional hypercube.
These were then shown to lead to a second order phase
transition at a critical point $k_c \sim \lambda / d$, 
summarized in the result of Eq.~(\ref{eq:kcd1}). 
Near this critical point it was found that all $\sim d^2$ lattice degrees
of freedom become massless (in the sense that all eigenvalues of the
quadratic fluctuation matrix have the same sign and approach zero),
suggesting a complete disappearance of the
conformal mode instability in the Euclidean theory at $d=\infty$,
in agreement with the naive conclusion from Eq.~(\ref{eq:prop}).

The second, and perhaps more ambitious, approach was based on a combined
strong coupling (small $k=1/8 \pi G$) and large $d$ expansion.
First, in the strong coupling expansion, it was found that the relevant
diagrams for the curvature correlation
function to a given order in $k$ can be identified with closed
surfaces, built out of parallel transport polygons, with
each polygon identified with the parallel transport of a test vector
around an elementary loop residing within the dual lattice.
We then argued that in the large $d$ limit it should be possible to
neglect surface self-intersections.
One then finds that such surfaces, based on their equivalent description
in terms of a two dimensional massless field theory, naturally
give rise to a logarithmic divergence of the correlation length
at the critical point at $k_c$, leading in this limit to the exact 
(and presumably universal) result of Eq.~(\ref{eq:xilog}).

The natural question then arises, and which is difficult
to ignore, of whether these large $d$ results
have any relevance for a physical four-dimensional world.
To the extent that the two cases are physically not too far apart,
one would be tempted to conclude that the dependence
of the correlation length $\xi$ on the gravitational coupling, as expressed
in Eq.~(\ref{eq:xilog}), and, conversely, the dependence of the
running gravitational coupling on $\xi$, as expressed
in Eq.~(\ref{eq:grun}), would suggest, for large $d$, finite but
{\it exponentially} small corrections to classical gravity,
at least in a scaling regime where the relevant distances involved
are much smaller then the macroscopic curvature scale,
$r \ll \xi \sim 1 / \sqrt{R} $,
but still much larger than the Planck scale, $r \gg l_P \sim \sqrt{G} $.
It is noteworthy that the quantum corrections computed here
are non-analytic at $r=0$, in spite of the fact that at short distances they
become rather small, and thus provide to some extent
a justification for the semi-classical picture of quantum gravity.
In terms of the parameters relevant for vacuum structure, the above-mentioned
non-perturbative curvature scale then corresponds
to a graviton vacuum condensate of order $\xi^{-1} \sim 10^{-30} eV$,
extraordinarily tiny compared to the QCD color condensate 
($\Lambda_{QCD}=220 MeV$) and the electro-weak Higgs condensate
($v=250 GeV$).
Furthermore, as has been stressed before, the quantum gravity theory, at least
in its present framework, does not and cannot
provide a value for the non-perturbative curvature scale $\xi$, 
which ultimately needs to be fixed by phenomenological input.
But, to the extent that this curvature scale clearly does not coincide with the
Planck scale (the cutoff scale), there is some room left for it to
take a very large, even cosmological, value.
The lattice gravity model in fact provides a clear case where the naive
identification of the curvature scale with the Planck scale
can be shown to be incorrect, due to the highly non-trivial renormalization
effects of strongly fluctuating quantum gravitational fields,
which cleverly arrange for the two scales to differ significantly in magnitude,
the more so as one approaches the critical point.

Finally, in the last section, we have attempted to make contact with known
results for the continuum theory above four dimension, and in 
particular those which
have some degree of relevance for the limit $d \rightarrow \infty$. 
Generally, and in analogy with the non-Abelian gauge theory case, 
it appears that the continuum theory does not seem to lead to the same level
of simplification as the regularized lattice gravity model discussed
in Sections 2 and 3
(and this in spite of their purported, but so far proven only for $d=3,4$,
equivalence in the lattice continuum limit).
Indeed in either case (gravity and gauge), the issue of perturbative 
non-renormalizability only gets worse with increasing dimension.
Ultimately we would tend to ascribe this state of affairs to the fact
that it appears quite challenging to perform the needed resummation of
the continuum theory with a bare cosmological constant
(as done explicitly only close to two dimensions),
perhaps an essential ingredient required to determine the true
non-perturbative, long distance behavior of quantum 
gravitation - even in infinitely many dimensions.

\vspace{20pt}

{\bf Acknowledgements}

The authors wish to thank the Theory Division at CERN
for warm hospitality and generous financial support during the
Summer of 2004, when this work was initiated, and during the
Summer of 2005 when this work was brought to completion.
The work of Ruth M. Williams was supported in part by the UK Particle
Physics and Astronomy Research Council.

\newpage

\vspace{20pt}

\appendix

\section*{Appendix}

\vskip 20pt
\newsection{Scalar Case and Random Walks}
\hspace*{\parindent}

The scalar field case is quite straightforward and is therefore 
worth reproducing here.
It relies on the well-known equivalence between the $\lambda \phi^4$ scalar
field theory and the Ising model, as far as their critical or long distance
behavior is concerned \cite{blz}.
The Ising partition function is given in any dimension by
\beq
Z ( \beta ) \; = \; \sum_{S_i = \pm 1} \, \exp \, 
\left [ \beta \sum_{<ij>} \, S_i \, S_j \right ] \;\; ,
\eeq
where $< \! ij \! >$ denotes a sum over nearest neighbors ($2d$ in 
$d$ dimensions, for a simple cubic lattice).
The corresponding scalar field theory is obtained by using a straightforward
Gaussian integral representation for the Ising statistical weight, which reads
\beq
\sum_{S_i = \pm 1} \, 
e^{ {\beta \over 2} \, \sum_{ij} \, S_i \, M_{ij} \, S_j }
\; = \;   (2 \pi)^{-{N \over 2}}  \left ( \det \beta M \right )^{-{1 \over 2}}
\, \int \prod_i d \, \phi_i \;
e^{ - \, {1 \over 2 \beta} \sum_{ij} \phi_i M_{ij}^{-1} \phi_j \, + \,
\sum_i \log ( 2 \cosh \phi_i ) } \;\; ,
\eeq
and then expanding the exponent in powers of the field $\phi$ and
its derivatives.
In either case the critical point is located where the renormalized
mass of the lowest excitation vanishes.
Returning to the Ising case, the spin susceptibility is then given by
\beq
\chi ( \beta )  \; = \; { 1 \over Z ( \beta ) } \, 
\sum_k \, \sum_{S_i = \pm 1} \, S_0 \, S_k \, \exp \, 
\left [ \beta \sum_{<ij>} \, S_i \, S_j \right ] \;\; ,
\eeq
and coincides with the spin correlation function
$ < \! S_0 \, S_k \! > $, summed over sites $k$.
Equivalently, it can be regarded as the Fourier transform of the
spin correlation function, evaluated at zero momentum.
It is convenient to re-write the formula for the partition function as
\beq
Z ( \beta ) \; = \; ( \cosh \beta )^N \, \sum_{S_i = \pm 1} \, 
\prod_{<ij>} \, \left [ 1 \, + \, t \, S_i \, S_j \right ] \;\; ,
\eeq
with $t = \tanh \beta $, $N$ the number of sites on the lattice,
and the product ranging over all links on the lattice. 
The expansion in $t$ has an obvious diagrammatic representation \cite{stan},
consisting in the case of $\chi (\beta) $ of open paths linking the site
$0$ to any site $k$, with each link appearing at most once
(but multiple times, if the expansion in $\beta$ is used instead). 
Write $\chi = \sum_n \chi_n t^n$, where $\chi_n$ is now the number of
open paths of length $n$ with fixed origin. 
We obtain a path of length $n+1$ by adding a 
link at its end, which can be done in $2d-1$ ways, giving 
$\chi_{n+1} \sim 2d \, \chi_n$, and so for large $d$ one
has \cite{fisher}
\beq
\chi ( \beta ) \; \propto \; \sum_{n=0}^{\infty} \, ( 2 d \, t )^n \; = \; 
{ 1 \over 1 - 2 d \, t } \;\; .
\eeq
Here use has been made of the fact that for large $d$ excluded volume effects
can be neglected, so that the factor $2d-1$ can simply be replaced
by $2d$.
Then from $\chi \sim 1/ (p^2 + m^2 ) \vert_{p=0} = \xi^2 $
(the spin correlation function evaluated at zero momentum)
one obtains $\xi \sim 1 / (t_c - t)^{\nu}$ with $t_c = 1/2d$ and $\nu =1/2$.

\vskip 20pt
\newsection{Vector case and q-coordinated Cayley Trees}
\hspace*{\parindent}

In the large $d$ limit dominant diagrams in the strong
coupling expansion of lattice gauge theories are represented
not by surfaces, but by trees made out of three-dimensional
cubes \cite{DPS,BDI}.
In the case of the plaquette-plaquette correlation function,
these are all the trees which can be constructed such that they
are anchored on the two given plaquettes.

The generating function for a q-coordinated Cayley tree \cite{trees}
(a Bethe lattice with $q$ links emanating from each vertex) is
given by \cite{DPS}
\beq
g(t) \; = \; { u \, ( 1 \, - \, 
{\textstyle {q \over 2} \displaystyle} \, u )
\over (1 - u )^2 } \;\; ,
\eeq
with parameters $u$ and $t$ related by
\beq
t \; = \; u \, ( 1 \, - \, u )^{q-2} \;\; ;
\label{eq:cons}
\eeq
$q=3$ corresponds to a trivalent or binary tree.
In the $SU(N)$ gauge case, one has $t = 2 d \beta^4$, 
with $\beta \approx 2 N / g^2 $
at strong coupling, and then the above is essentially the same as
the free energy of the gauge theory (up to various inessential constants).
Also, in the gauge case $q=6$ (since a cube has six faces) and a new
cube can be attached to any of the six faces of the original
cube (again ignoring excluded volume effects at large $d$), 
thus creating a continuous tree made out of cubes. 
The free energy is then equal to a sum over all possible trees
of arbitrary length, giving rise to hydra-like configurations as
viewed from the diagrammatic perspective of the strong coupling
expansion.

In particular, the plaquette-plaquette correlation function
is obtained from the second derivative of the above generating
function $g(t)$ with respect to the coupling $\beta$.
It can be represented as the sum of all trees of
arbitrary shape (with coordination $q=6$), but now
with two fixed endpoints.
Extending the analysis to general $q$, one can show that
in fact the key result is in fact independent of $q$ for $q>2$.
The relevant singularity in the second derivative of the
free energy $g(t)$ corresponds to $u_c = 1 / (q-1)$.
Expanding Eq.~(\ref{eq:cons}) in the vicinity of this point one finds
\beq
t \; = \; t_c \, - \, \half \, (q-1)^{4-q} \, (q-2)^{q-3} \,
\left ( u \, - \, {1 \over q-1} \right )^2 + \dots
\eeq
i.e. the linear term vanishes.
In the above expression $t_c$ is the critical point,
\beq
t_c \; = \; { ( q - 2)^{q-2} \over ( q-1)^{q-1} } \;\; .
\eeq
Thus $t_c-t \sim (u - u_c)^2 $ for any $q>2$.
First and second derivatives of the free energy $g(t)$ with
respect to $t$ can then be calculated via
\bea
{d \, g \over d \, u} \; & = & \; { (q-1) \, u - 1 \over (u-1)^3 }
\nonumber \\
{d \, g \over d \, t } \; & = & \; 
{d \, g \over d \, u } \, {d \, u \over d \, t } \; = \; 
{ 1 \over (1-u)^q }
\nonumber \\
{d^2 g \over d \, t^2 } \; & = & \; 
{ q \, (1-u)^{2 - 2 q} \over 1 + (1-q) \, u } \;\; ,
\eea
which for any $q>2$ behaves in the limit $t \rightarrow t_c $ as
\beq
{d^2 g \over d \, t^2 } \; \sim \; 
{ q \, ( q - 2)^{1-3q \over 2} \over \sqrt{2} \, 
( q-1)^{2-3q \over 2} }
\; { 1 \over \sqrt{ t_c - t } } \sim \xi^2 \;\; .
\eeq
Here use has been made of the fact that the second derivative
of the free energy brings down two plaquette terms, giving
the plaquette-plaquette correlation function, summed over
both plaquette coordinates, and which is therefore
equivalent to the Fourier transform of the plaquette-plaquette
correlation at zero momentum. 
Thus one obtains the momentum space plaquette-plaquette correlation at
zero momentum, or $1/(p^2 + m^2) \vert_{p=0}$, with $m = \xi^{-1}$,
and this then gives $\xi \sim 1 / (t_c - t)^{1/4}$ and thus $\nu = 1/4$ for
any $q > 2$.
It is further observed in \cite{DPS} that the second order phase transition
of the gauge theory described by $g(t)$ bears a striking similarity
to the condensation of branched polymers, with the polymer chain
built out of (trees of) three-dimensional cubes.

\vfill

\newpage


\vskip 40pt


\newpage





\begin{thebibliography}{99}



\bibitem{feylec}
  R.~P.~Feynman, {\sl `Lectures on Gravitation'}, 1962-1963,
  edited by F.~B.~Morinigo and W.~G.~Wagner, California Institute
  of Technology (Pasadena, 1971); {\sl Quantum Theory of Gravitation},
  {\it Acta Phys.\ Pol.}, vol.~XXIV, 697-722 (1963).


\bibitem{dewittlec}
  B.~S.~DeWitt, {\sl `Dynamical Theory of Groups and Fields'}, 
  {\it Les Houches} Lectures 1963, (Gordon and Breach, New York, 1965).


\bibitem {thooft}
  G.~'t Hooft,
  {\it Nucl.\ Phys.} {\bf B 62}, 444 (1973);
  G.~'t Hooft and M.~Veltman, 
  {\it Ann. Inst. Poincar\'e} {\bf 20} 69 (1974).


\bibitem{deser}
  S.~Deser and P.~van Nieuwenhuizen, {\it Phys. Rev.}{\bf D10} 401,410 (1974);
  S.~Deser, H.~S.~Tsao and P.~van Nieuwenhuizen,
  {\it Phys.\ Rev.} {\bf D 10}, 3337 (1974);
  P.~Van Nieuwenhuizen,
  {\it Annals Phys.} {\bf 104}, 197 (1977).




\bibitem{sagnotti}
  M.~H.~Goroff and A.~Sagnotti,
  {\it Phys.\ Lett.} {\bf B 160}, 81 (1985);
  {\it Nucl.\ Phys.} {\bf B 266}, 709 (1986).


\bibitem{vandeven}
  A.~E.~M.~van de Ven,
  {\it Nucl.\ Phys.} {\bf B 378}, 309 (1992).





\bibitem{wilson}
  K.~G.~Wilson, {\it Rev.\ Mod.\ Phys.} {\bf 47} 773 (1975);
  {\bf 55} 583 (1983);
  K.~G.~Wilson and M.~Fisher, {\it Phys.\ Rev.\ Lett.} {\bf 28} 240 (1972).






\bibitem{wilseps}
  K.~G.~Wilson, {\it Phys.\ Rev.} {\bf D 7}, 2911 (1973).


\bibitem {parisi}
  G.~Parisi, {\it Nucl.\ Phys.} {\bf B100} 368 (1975), {\it ibid.} 
  {\bf 254} 58 (1985); {\sl `On Non-renormalizable Interactions'}, 
  in {\it New Development in Quantum Field Theory and Statistical Mechanics},
  (Cargese 1976, M.~Levy and P.~Mitter eds. , Plenum Press, New York 1977).


\bibitem{sym}
  K.~Symanzik, {\it Comm.\ Math.\ Phys.} {\bf 45} 79-98 (1975).


\bibitem{nonren-wein}
  S.~Weinberg, {\it Phys.\ Rev.} {\bf D56} 2303 (1997).


\bibitem {hawking}
  S.~W.~Hawking, in {\sl `General Relativity - An Einstein Centenary
  Survey'}, edited by S.~W.~Hawking and W.~Israel,
  (Cambridge University Press, 1979);
  S.~W.~Hawking and T.~Hertog,
  {\it Phys.\ Rev.} {\bf D 65}, 103515 (2002).


\bibitem{harhawk}
  J.~B.~Hartle and S.~W.~Hawking,
  {\it Phys.\ Rev.} {\bf D 28}, 2960 (1983).


\bibitem{harcosm}
  J.~B.~Hartle,
  lectures delivered at the {\it Theoretical Advanced Study Institute}
  in Elementary Particle Physics, Yale University, 1985, vol. 2, p. 471-566. 


\bibitem {cosm}
  H.~W.~Hamber and R.~M.~Williams, {\it Phys.\ Rev.} {\bf D 72} 44026 (2005);
  gr-qc/0506137.





\bibitem {regge}
  T.~Regge, {\it Nuovo Cimento} {\bf 19} 558 (1961).


\bibitem {wheeler}
 J.~A.~Wheeler, {\sl `Geometrodynamics and the Issue of the Final State'},
 in {\it Relativity, Groups and Topology}, Les Houches 1963,
 edited by C.~De Witt and B.~S.~De Witt (Gordon and Breach, New York, 1964).


\bibitem {rowi}
  M.~Ro\u cek and R.~M.~Williams, {\it Phys.\ Lett.} {\bf 104B} 31 (1981);
  {\it Z.\ Phys.} {\bf C21} 371 (1984).


\bibitem {lee}
  T.~D.~Lee, in {\sl `Discrete Mechanics'}, 
  1983 Erice International School of Subnuclear Physics,
  vol. 21 (Plenum Press, New York 1985), and references therein.


\bibitem {hw84}
  H.~W.~Hamber and R.~M.~Williams, {\it Nucl.\ Phys.} {\bf B248} 392 (1984);
  {\bf B260} 747 (1985); 
  {\it Phys.\ Lett.} {\bf 157B} 368 (1985);
  {\it Nucl.\ Phys.} {\bf B267} 482 (1986); {\bf B269} 712 (1986).


\bibitem {hartle}
  J.~B.~Hartle, {\it J.\ Math.\ Phys.} {\bf 26} 804 (1985);
  {\bf 27} 287 (1985); {\bf 30} 452 (1989).


\bibitem {lesh} 
  H.~W.~Hamber, in {\it Critical Phenomena, Random Systems, Gauge Theories}, 
  1984 {\it Les Houches} Summer School, Session XLIII, 
  (North Holland, Amsterdam, 1986).


\bibitem {monte}
  B.~Berg,  {\it Phys.\ Rev.\ Lett.} {\bf 55} 904 (1985);
  {\it Phys.\ Lett.} {\bf B176} 39 (1986);
  J.~Riedler, W.~Beirl, E.~Bittner, A.~Hauke, P.~Homolka and H.~Markum,
  {\it Class.\ Quant.\ Grav.} {\bf 16}, 1163 (1999), and references
  therein.


\bibitem {critical}
  H.~W.~Hamber, {\it Phys.\ Rev.} {\bf D45} 507 (1992);
  {\it Phys.\ Rev.} {\bf D61} 124008 (2000); gr-qc/9809090.








\bibitem {epsilon}
  R.~Gastmans, R.~Kallosh and C.~Truffin, 
  {\it Nucl.\ Phys.} {\bf B133} 417 (1978);  
  S.~M.~Christensen and M.~J.~Duff, {\it Phys.\ Lett.} {\bf B79} 213 (1978).


\bibitem{epsilon1}
  S.~Weinberg, in {\sl `General Relativity - An Einstein Centenary
  Survey'}, edited by S.~W.~Hawking and W.~Israel,
  (Cambridge University Press, 1979).


\bibitem {epsilon2}
  H.~Kawai and M.~Ninomiya,  {\it Nucl.\ Phys.} {\bf B336} 115 (1990);  
  H.~Kawai, Y.~Kitazawa and M.~Ninomiya,  {\it Nucl.\ Phys.} 
  {\bf B393} 280 (1993) and {\bf B404} 684 (1993);  
  Y.~Kitazawa and M.~Ninomiya, {\it Phys.\ Rev.} {\bf D55} 2076 (1997).


\bibitem {epsilon3}
  T.~Aida and Y.~Kitazawa, {\it Nucl.\ Phys.} {\bf B491} 427 (1997).






\bibitem{litim}
  D.~Litim, CERN-TH-2003-299, {\it Phys.\ Rev.\ Lett.} {\bf 92} 201301 (2004). 


\bibitem{reuter}
  O.~Lauscher and M.~Reuter, {\it Class.\ Quant.\ Grav.} {\bf 19} 483 (2002);  
  M.~Reuter and F.~Saueressig, {\it Phys.\ Rev.} {\bf D65} 065016 (2002).






\bibitem{englert}
  F.~Englert, {\it Phys.\ Rev.} {\bf 129}, 567 (1963).


\bibitem{fisher}
  M.~E.~Fisher and D.~S.~Gaunt, {\it Phys.\ Rev.} {\bf 133}, A224 (1964).


\bibitem{abe}
  R.~Abe, {\it Progr.\ Theor.\ Phys.} {\bf 47}, 62 (1972).






\bibitem{stanley}
  H.~E.~Stanley, {\it Phys.\ Rev.} {\bf 176}, 718 (1968);
  E.~Brezin and D.~J.~Wallace, {\it Phys.\ Rev.} {\bf B 7}, 1967 (1973);
  K.~G.~Wilson, {\it Phys.\ Rev.} {\bf D 7}, 2911 (1973).


\bibitem{zinn}
  J.~Zinn-Justin, {\sl `Vector Models in the Large-N Limit'},
  hep-th/9810198, and references therein;
  See also {\sl `Quantum Field Theory and Critical Phenomena'},
  (Oxford University Press, 3-rd edition, 1996).


\bibitem{thooft1}
  G.~'t Hooft,
  {\it Nucl.\ Phys.} {\bf B 72}, 461 (1974).


\bibitem{thooft2}
  G.~'t Hooft, {\sl 'Large N'}, arXiv:hep-th/0204069, and references
  therein.







\bibitem {cms1}
  J.~Cheeger, W.~M\"uller and R.~Schrader, 
  in {\sl `Unified Theories Of Elementary Particles'},  
  {\it Heisenberg Symposium}, M\"unchen 1981, p. 176-188, 
  (Springer, New York, 1982).


\bibitem {det}
  H.~W.~Hamber and R.~M.~Williams, {\it Phys.\ Rev.} {\bf D59} 064014 (1999);
  {\bf D70}, 124007 (2004), and references therein.


\bibitem{coxeter}
  H.~Coxeter, {\sl `Regular Polytopes'}, Methuen and Co. Ltd., London, 1948;
  {\sl `Regular Complex Polytopes'}, Cambridge University press, 1974.


\bibitem {hw3d}
H.~W.~Hamber and R.~M.~Williams, {\it Phys.\ Rev.} {\bf D47} 510 (1993).







\bibitem {dewitt}
  B.~DeWitt, {\it Phys.\ Rev.} {\bf 160} 1113 (1967);
  in {\sl `General Relativity - An Einstein Centenary Survey'},
  edited by S.~W.~Hawking and W.~Israel, Cambridge University Press (1979).


\bibitem {misner}
  C.~W.~Misner, {\it Rev. Mod. Phys.} {\bf 29} 497 (1957);
  L.~D.~Faddeev and V.~N.~Popov, {\it Sov.\ Phys.\ Usp.} 
  {\bf 16} 777-788 (1974);
  {\it Usp.\ Fiz.\ Nauk.} {\bf 111} 427-450 (1973).






\bibitem{dombgreen}
  See for example, 
  C.~Domb and H.~S.~Green, 
  {\sl `Phase Transitions and Critical Phenomena'},
  Vol. 3 (London 1976);
  C.~Domb and J.~L.~Lebowitz, 
  {\sl `Phase Transitions and Critical Phenomena'},
  Vol. 13 (London 1989).


\bibitem{pade}
  G.~A.~Jr.~Baker and P.~Graves-Morris, {\sl `Pad\'e Approximants'}, 
  Cambridge University Press, New York, 1996. 





\bibitem{pearson}
  C.~Itzykson, R.~B.~Pearson and J.~B.~Zuber,
  {\it Nucl.\ Phys.} {\bf B 220}, 415 (1983).


\bibitem {qftexp}
  J.~C.~Le Guillou and J.~Zinn-Justin, {\it Phys.\ Rev.} {\bf B21} 3976 (1980);
  {\it J.\ Phys.} (France) {\bf 50} 1365 (1989).


\bibitem{guida}
  R.~Guida and J.~Zinn-Justin,
  cond-mat/9803240.


\bibitem{campo}
  M.~Campostrini, A.~Pelissetto, P.~Rossi and E.~Vicari,
  {\it Phys.\ Rev.} {\bf E 60}, 3526 (1999).





\bibitem{caselle}
  M.~Caselle, A.~D'Adda and L.~Magnea,
  {\it Phys.\ Lett.} {\bf B 232}, 457 (1989).


\bibitem{froh}
  J.~Fr\"ohlich, {\sl `Regge Calculus and Discretized Gravitational Functional
  Integrals'}, I.~H.~E.~S.~preprint 1981 (unpublished);
  {\it Non-Perturbative Quantum Field Theory: Mathematical Aspects and
  Applications}, Selected Papers (World Scientific, Singapore, 1992),
  pp. 523-545. 



\bibitem{haar}
  N.~J.~Vilenkin and A.~U.~Klymik,
  {\sl `Representation of Lie groups and Special
  Functions'}, Volume 2, Mathematics and its Applications, 
  Kluwer Academic Publisher, 1993.


\bibitem{bianchi}
  H.~W.~Hamber and G.~Kagel, {\it Class.\ Quant.\ Grav.} {\bf 21}, 5915 (2004),
  and references therein.


\bibitem{wilson-lgt}
  K.~G.~Wilson,
  {\it Phys.\ Rev.} {\bf D 10}, 2445 (1974);
  {\sl `Quarks And Strings On A Lattice'},
  in {\it New Phenomena In Subnuclear Physics}, Erice 1975 
  (Plenum Press, New York, 1977).


\bibitem{frampton}
  See, for example, P.~H.~Frampton, {\sl `Gauge Field Theories'}, ch. 6 
  (John Wiley \& Sons, New York, 2000).



\bibitem{moda}
  G.~Modanese,
  {\it Phys.\ Rev.} {\bf D 47}, 502 (1993);
  {\it Phys.\ Rev.} {\bf D 49}, 6534 (1994).


\bibitem {lines}
  H.~W.~Hamber and R.~M.~Williams, {\it Nucl.\ Phys.} {\bf B435} 361 (1995).

\bibitem {peskin}
  See, for example, M.~E.~Peskin and D.~V.~Schroeder, 
  {\sl `An Introduction to Quantum Field Theory'}, sec. 22.1 
  (Addison Wesley, Reading, Massachusetts, 1995).


\bibitem {feypath}
  R.~P.~Feynman and A.~Hibbs, {\sl `Quantum Mechanics and Path Integrals'}
  (McGraw Hill, New York, 1965).



\bibitem{symrw}
  K.~Symanzik, in {\sl `Euclidean Quantum Field Theory'},
  Proceeding of the International School of Physics ``{\it Enrico Fermi}'',
  Varenna, Session XLV, edited by R.~Jost (Academic Press, New York, 1969).


\bibitem{haus}
  G.~Parisi,
  {\it Phys.\ Lett.} {\bf B 81}, 357 (1979), and references therein.


\bibitem{gross-dtrs}
  D.~J.~Gross,
  {\it Phys.\ Lett.} {\bf B 138}, 185 (1984).




\bibitem{phi4}
  M.~Aizenman, {\it Phys.\ Rev.\ Lett.} {\bf 47} 1 (1981); 
  {\it Commun.\ Math.\ Phys.} {\bf 86} 1 (1982); {\bf 97} 91 (1985); \\
  J.~Fr\"ohlich, {\it Nucl.\ Phys.} {\bf B200} 281 (1982).



\bibitem {ttmodes}
  H.~W.~Hamber and R.~M.~Williams, 
  {\it Phys.\ Rev.} {\bf D 70}, 124007 (2004) [hep-th/0407039].


\bibitem {hw2d}
  H.~W.~Hamber and R.~M.~Williams, {\it Nucl.\ Phys.} {\bf B267} 482 (1986).



\bibitem{DPS}
  J.~M.~Drouffe, G.~Parisi and N.~Sourlas,
  {\it Nucl.\ Phys.} {\bf B 161}, 397 (1979).









\bibitem{caswell}
  W.~E.~Caswell,
  {\it Phys.\ Rev.\ Lett.} {\bf 33}, 244 (1974);
  D.~R.~T.~Jones,
  {\it Nucl.\ Phys.} {\bf B 75}, 531 (1974).






\bibitem {hdqg}
  E.~S.~Fradkin and A.~A.~Tseytlin, {\it Phys.\ Lett.} {\bf 104B} 377 (1981)
  and {\bf 106B} 63 (1981); {\it Nucl.\ Phys.} {\bf B201} 469 (1982);
  I.~G.~Avramidy and A.~O.~Barvinsky,{\it Phys.\ Lett.} {\bf 159B} 269 (1985).


\bibitem{deser04}
  S.~Deser,
  gr-qc/0411026.
  See also S.~Deser, {\sl The Many Dimensions of Dimension}, 
  physics/0402105.


\bibitem{deser3d}
  S.~Deser, R.~Jackiw and S.~Templeton,
  {\it Annals Phys.} {\bf 140}, 372 (1982); 
  {\it ibid.} {\bf 185}, 406 (1988); \\
  see also S.~Deser, R.~Jackiw and G.~'t Hooft,
  {\it Annals Phys.} {\bf 152}, 220 (1984).






\bibitem {sigma}
  E.~Brezin and J.~Zinn-Justin, {\it Phys.\ Rev.\ Lett.} {\bf 36} 691 (1976); 
  {\it Phys.\ Rev.} {\bf D14} 2615 (1976); 
  {\it Phys.\ Rev.} {\bf B14}, 3110 (1976);
  E.~Brezin and S.~Hikami, LPTENS-96-64 (Dec 1996) [cond-mat/9612016].


\bibitem{grossneveu}
  D.~J.~Gross and A.~Neveu,
  {\it Phys.\ Rev.} {\bf D 10}, 3235 (1974).


\bibitem{hikami}
  S.~Hikami and E.~Brezin,
  {\it J.\ Phys.} {\bf A 11}, 1141 (1978).


\bibitem{kleinert}
  H.~Kleinert,
  {\it Phys.\ Lett.} {\bf A 264}, 357 (2000).


\bibitem{pot}
  H.~W.~Hamber and S.~Liu,
  {\it Phys.\ Lett.} {\bf B 357}, 51 (1995).

\bibitem{fpgrav}
  L.~D.~Faddeev and V.~N.~Popov, {\it Sov.\ Phys.\ Usp.} 
  {\bf 16} 777-788 (1974);
  {\it Usp.\ Fiz.\ Nauk.} {\bf 111} 427-450 (1973).
 
\bibitem {capper}
  D.~M.~Capper, G.~Leibbrandt and M.~Ramon Medrano,
  {\it Phys.\ Rev.} {\bf D8} 4320 (1973).

\bibitem {fp}
  L.~D.~Faddeev and V.~N.~Popov, {\it Phys.\ Lett.} {\bf 25B} 29 (1967).

\bibitem {ft}
  E.~S.~Fradkin and I.~V.~Tyutin,
  {\it Phys.\ Rev.} {\bf D2} 2841 (1970).







\bibitem{strom1}
  A.~Strominger,
  {\it Phys.\ Rev.} {\bf D 24}, 3082 (1981).


\bibitem{strom2}
  A.~Strominger,
  Print-82-0798 (IAS,Princeton), talk presented at the 
  Int. Symp. on {\it Gauge Theory and Gravitation},
  Nara, Japan, Aug 20-24, 1982.




\bibitem{bb}
  N.~E.~J.~Bjerrum-Bohr,
  {\it Nucl.\ Phys.} {\bf B 684}, 209 (2004).








\bibitem{blz}
  E.~Brezin, J.~C.~Le Guillou and J.~Zinn-Justin,
  {\sl `Field Theoretical Approach To Critical Phenomena'},
  in ``Phase Transitions and Critical Phenomena'', 
  Vol. 6, edited by C. Domb and M. Green, London 1976, 125-247. 


\bibitem{stan}
  H.~E.~Stanley, 
  {\sl `Introduction to Phase Transitions and Critical Phenomena'},
  ch. 9 (Oxford University Press, New York and Oxford, 1971).




\bibitem{BDI}
  R.~Balian, J.~M.~Drouffe and C.~Itzykson,
  {\it Phys.\ Rev.} {\bf D 10}, 3376 (1974);
  {\it Phys.\ Rev.} {\bf D 11}, 2098 (1975);
  {\it Phys.\ Rev.} {\bf D 11}, 2104 (1975);
  [Erratum-ibid.\ {\bf D 19}, 2514 (1979)].




\bibitem{trees}
  F.~Harary, {\sl `Graph Theory'}, ch. 4, 
  (Addison-Wesley, Reading, Mass. ,1994). 




















\end{thebibliography}
\end{document}